\documentclass[10pt,conference]{IEEEtran}
\usepackage{cite}
\usepackage{amsmath,amssymb,amsfonts}
\usepackage{algorithmic}
\usepackage{graphicx}
\usepackage{textcomp}
\usepackage{xcolor}
\usepackage[hyphens]{url}

\usepackage{algorithm}
\usepackage{subcaption}
\usepackage{authblk}

\def\BibTeX{{\rm B\kern-.05em{\sc i\kern-.025em b}\kern-.08em
    T\kern-.1667em\lower.7ex\hbox{E}\kern-.125emX}}

\title{Wide Quantum Circuit Optimization with Topology Aware Synthesis}
\author[1]{Mathias Weiden}
\author[1]{Justin Kalloor}
\author[1]{John Kubiatowicz}
\author[2]{Ed Younis}
\author[2]{Costin Iancu}
\affil[1]{Department of Electrical Engineering and Computer Science, University of California, Berkeley}
\affil[ ]{\textit{\{mtweiden, jkalloor3, kubitron\}@cs.berkeley.edu}}
\affil[2]{Computational Research Division, Lawrence Berkeley National Laboratory}
\affil[ ]{\textit{\{edyounis, cciancu\}@lbl.gov}}

\title{Wide Quantum Circuit Optimization with Topology Aware Synthesis} 

\begin{document}
\maketitle
\thispagestyle{plain}
\pagestyle{plain}


\begin{abstract}
Unitary synthesis is an optimization technique that can achieve optimal multi-qubit gate counts while mapping quantum circuits to restrictive qubit topologies. 
Because synthesis algorithms are limited in scalability by their exponentially growing run time and memory requirements, application to circuits wider than 5 qubits requires divide-and-conquer partitioning of circuits into smaller components.
In this work, we will explore methods to reduce the depth (program run time) and multi-qubit gate instruction count of wide (16-100 qubit) mapped quantum circuits optimized with synthesis. 
Reducing circuit depth and gate count directly impacts program performance and the likelihood of successful execution for quantum circuits on parallel quantum machines.

We present TopAS, a topology aware synthesis tool built with the \emph{BQSKit} framework that preconditions quantum circuits before mapping. 
Partitioned subcircuits are optimized and fitted to sparse qubit subtopologies in a way that balances the often opposing demands of synthesis and mapping algorithms.
This technique can be used to reduce the depth and gate count of wide quantum circuits mapped to the sparse qubit topologies of Google and IBM. 
Compared to large scale synthesis algorithms which focus on optimizing quantum circuits after mapping, TopAS is able to reduce depth by an average of 35.2\% and CNOT gate count an average of 11.5\% when targeting a 2D mesh topology.
When compared with traditional quantum compilers using peephole optimization and mapping algorithms from the Qiskit or $t|ket\rangle$ toolkits, our approach is able to provide significant improvements in performance, reducing CNOT counts by 30.3\% and depth by 38.2\% on average.
\end{abstract}

\section{Introduction}
\label{section:introduction}
Modern quantum machines are subject to high levels of environmental noise, are difficult to control, and consist of only tens of qubits.
Quantum error correction and mitigation are widely regarded as vital components in the development of practical quantum computers \cite{preskill}.
However, because quantum error correction is not yet realistically implementable, it is essential that quantum programs be optimized so that run times and operation counts are minimized, as each operation adds additional error into the system and longer run times increase the probability that qubits will decohere.
Operations that involve multiple qubits, such as the \emph{CNOT}, \emph{iSWAP}, \emph{Toffoli}, and \emph{CZ} gates, are typically far more error prone and expensive to implement compared to single qubit rotation gates \cite{gambetta_cramming_2019}.
For this reason, multi-qubit gate counts can be used as a metric for the error introduced into quantum programs.

Optimization techniques that aim to decrease the multi-qubit gate count of quantum circuits are becoming increasingly used and studied \cite{li2019tackling, younis_qfast_2020, shafaei_optimization_2013}.
Unitary synthesis is a method of decomposing a quantum circuit into a simpler quantum circuit that fits some specified gate set. 
Synthesis seeks to produce shorter circuits with reduced depth and gate count in order to improve the likelihood that circuits are executed correctly.

The runtime of unitary synthesis algorithms scales exponentially with the number of qubits $n$. 
For current state of the art unitary synthesis algorithms such as the QSearch/LEAP algorithm, only circuits up to 5 qubits are realistically decomposable.
Larger circuits must thus be partitioned into smaller subcircuits.
After synthesizing each subcircuit independently, subcircuits in the original circuit are replaced with their synthesized versions.

\begin{figure}
    \begin{subfigure}[t]{0.23\textwidth}
        \centering
        \includegraphics[width=.6\textwidth]{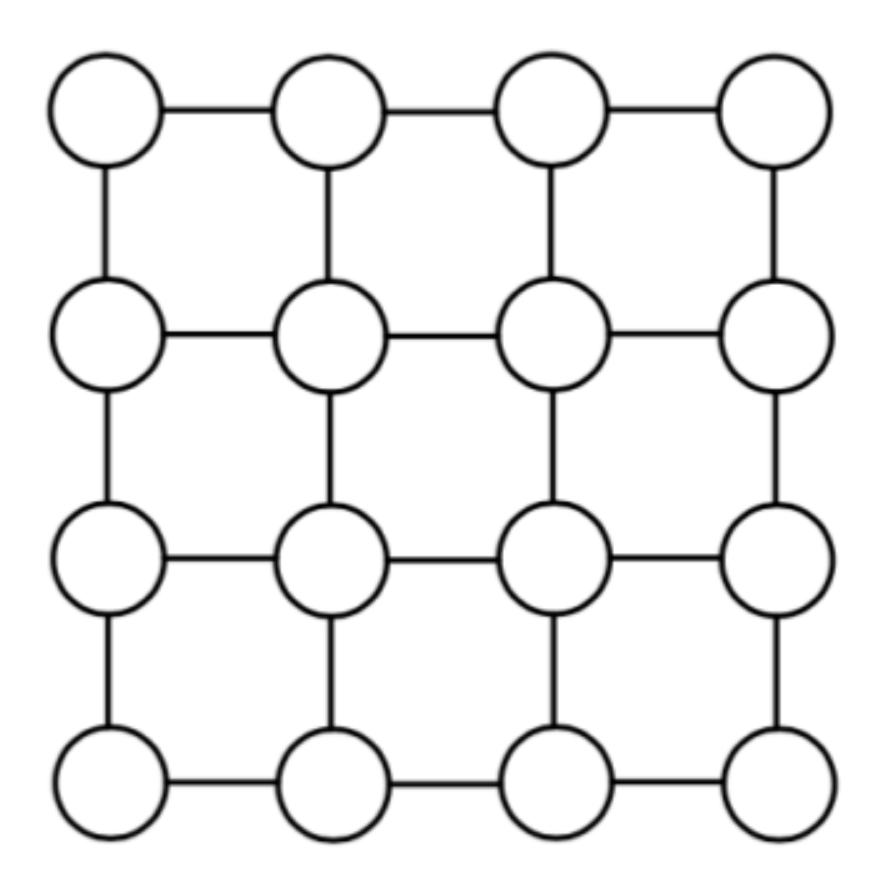}
        \caption{Mesh}
        \label{fig:mesh_topology}
    \end{subfigure}
    \begin{subfigure}[t]{0.23\textwidth}
        \centering
        \includegraphics[width=.6\textwidth]{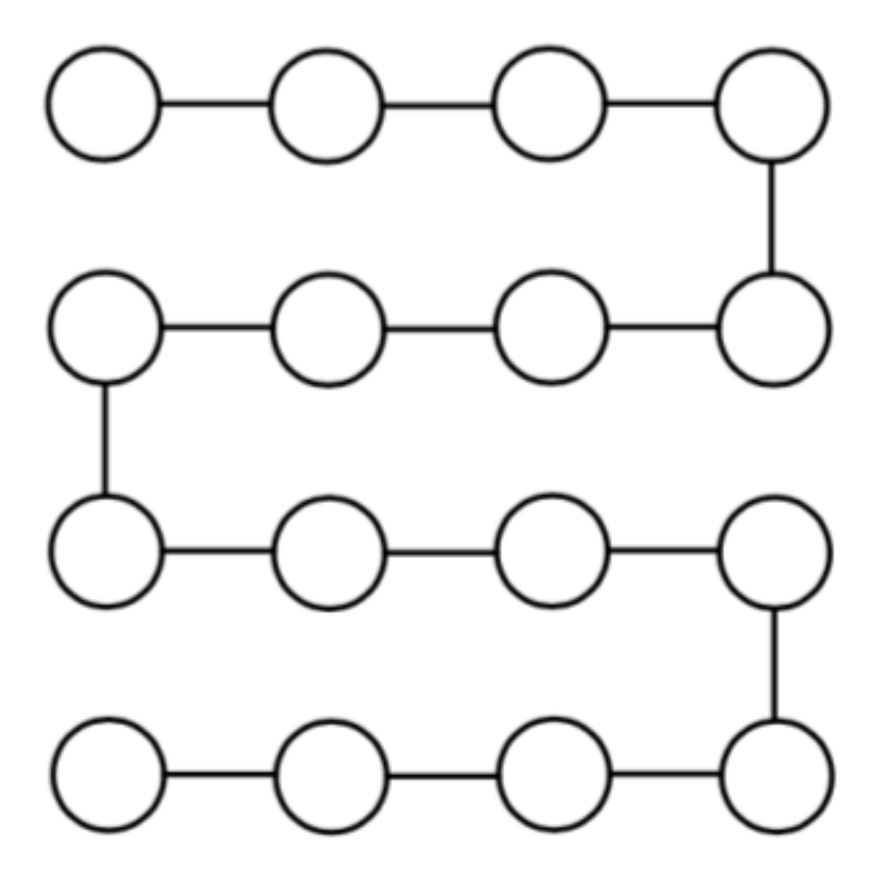}
        \caption{Linear}
        \label{fig:linear_topology}
    \end{subfigure}
    \begin{subfigure}[b]{0.5\textwidth}
        \centering
        \includegraphics[width=.5\textwidth]{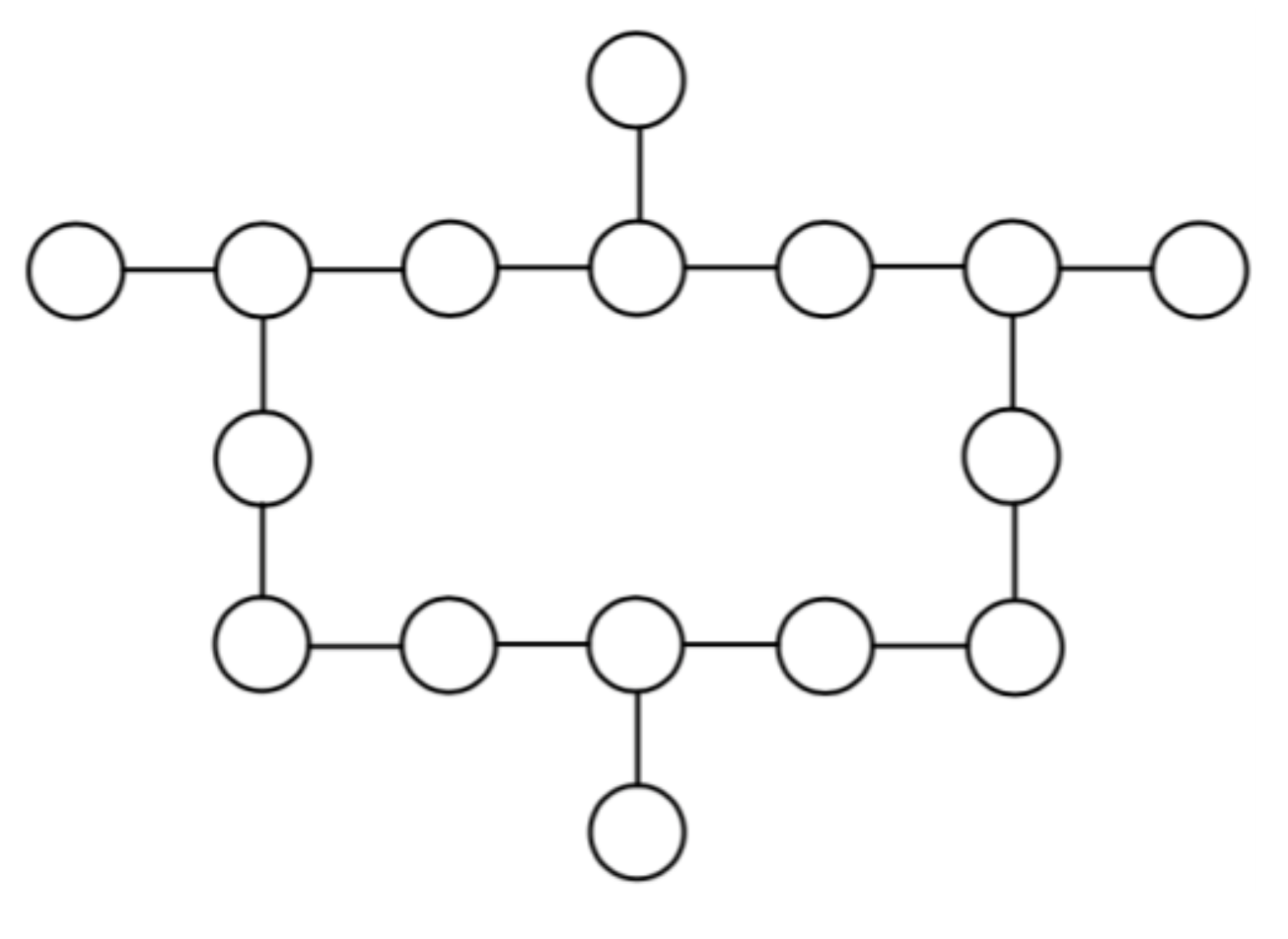}
        \caption{Falcon}
        \label{fig:falcon_topology}
    \end{subfigure}
    \caption{Example Physical Qubit Topologies. The 2D mesh topology is based off quantum machines provided by Google, while the falcon topology follows IBM's heavy hexagonal connectivity scheme.}
    \label{fig:physical_topologies}
\end{figure}

Other than sensitivity to environmental noise, machines in the Noisy Intermediate Scale Quantum (NISQ) era are defined by their limited sizes and connectivities (which physical qubits are allowed to interact). 
The connectivity of qubits in machines using superconducting qubits is especially restrictive.
Examples of realistic and popular superconducting qubit physical topologies are illustrated in Figure \ref{fig:physical_topologies}.

Each vertex in a physical qubit topology graph represents a hardware qubit.
The set of edges between these hardware qubits is the set of supported multi-qubit interactions for that quantum machine.
Quantum algorithms are typically designed assuming that each qubit is able to interact directly with all other qubits: logical circuits assume all to all connectivity.
In order for a quantum processor to execute these densely connected logical circuits, a qubit mapping algorithm must insert SWAP gates routing all interactions along edges in the physical topology.

As quantum circuits become wider and physical topologies remain sparse, the number of routing gates inserted by these mapping algorithms grows quickly.
It is therefore desirable to approximate and transform wide quantum circuits using unitary synthesis so that the total number of multi-qubit gates and depth of quantum circuits is reduced. 

In this paper, we present TopAS, a qubit topology aware synthesis tool that applies unitary synthesis to wide quantum circuits to reduce multi-qubit gate counts and circuit depth. 
TopAS first partitions a logical quantum circuit, synthesizes each partitioned subcircuit independently, reassembles the optimized logical circuit, then uses a mapping algorithm to ensure that all operations are legal according to some specified qubit topology.
When targeting the mesh physical topology, TopAS is able to produce circuits with an average of 30.1\% fewer CNOT gates than \emph{Qiskit} \cite{qiskit} and $t|ket\rangle$ \cite{sivarajah2020t}, and 11.5\% fewer CNOT gates than other large scale synthesis techniques such as QGo \cite{qgo}.
TopAS improves upon the QGo algorithm by optimizing before mapping and by preconditioning circuit partitions so that the fully mapped results are less deep and require fewer multi-qubit gates.

The remainder of this paper is structured as follows: Section \ref{section:background} discusses the general processes of unitary optimization, quantum circuit partitioning, and compares tools that apply synthesis before and after mapping.
Section \ref{section:topas} presents the design choices made for the TopAS tool. 
Section \ref{section:data} compares the TopAS tool to other optimization and mapping techniques.
Finally, Section \ref{section:discussion} provides commentary and discussion about advancements that could further improve the performance of wide quantum circuit synthesis tools.

\section{Background}
\label{section:background}
\subsection{Quantum Computing Basics}
The fundamental unit of information in a quantum computer is the qubit, which can be represented as a vector of the form
\[
    | \psi \rangle = \alpha \begin{bmatrix} 1 \\ 0 \end{bmatrix} + \beta \begin{bmatrix} 0 \\ 1 \end{bmatrix}
\]
where $\alpha$ and $\beta$ are complex numbers such that $|\alpha|^2 + |\beta|^2 = 1$ and $[ 1 \; 0 ]^T$ and $[ 0 \; 1 ]^T$ are orthonormal basis vectors representing two distinct quantum states.
The state of a quantum system with $n$ qubits in states $|\psi_1\rangle, |\psi_2\rangle, \dots |\psi_n\rangle$ lies in a $2^n \times 2^n$ Hilbert space, and can be represented by  the tensor product $|\psi_1\rangle \otimes |\psi_2\rangle \otimes \dots \otimes |\psi_n\rangle$.
This quantum state $|\psi\rangle$ can be evolved by use of $2^n \times 2^n$ unitary operators\cite{nielsen2002quantum}.

\begin{figure}
    \centering
    \includegraphics[width=6cm]{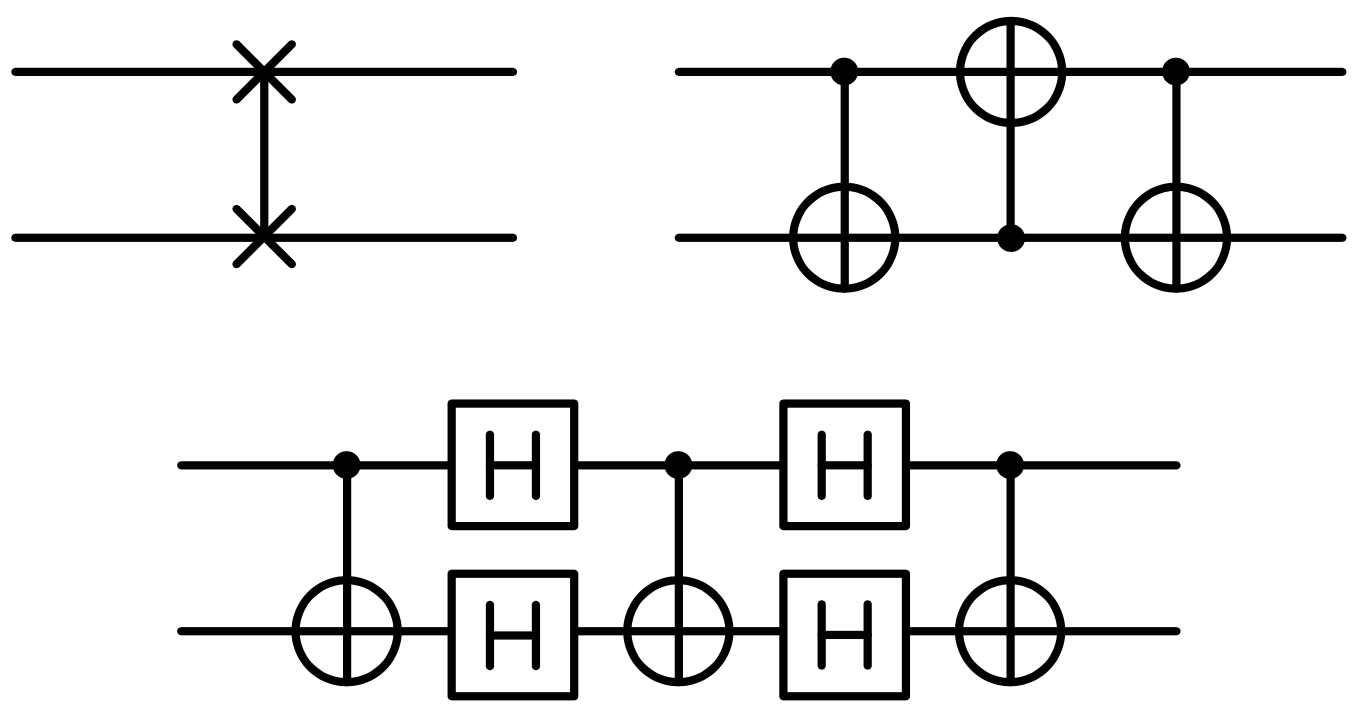}
    \caption{How to implement a SWAP operation using a SWAP gate, 3 CNOTs, and 3 CNOTs and 4 Hadamards.}
    \label{fig:swap_gate}
\end{figure}

The quantum circuit model represents this unitary as a series of quantum gates \cite{deutsch1989quantum}.
Single qubit and multi-qubit gates act on qubits which are drawn as horizontal wires (see Figure \ref{fig:q_circuit}). 
The number of wires or qubits $n$ is called the circuit width, while the critical path length or depth of the circuit is $T$.
The depth is typically drawn as the x-axis of the circuit. 
For the scope of this paper, a universal gate set of \{U3, CNOT\} is assumed. 
Note that SWAP gates can trivially be decomposed into 3 CNOTs as shown in Figure \ref{fig:swap_gate}.

\subsection{Quantum Circuit Synthesis}
Given a target $2^n \times 2^n$ unitary $U$ and an error threshold $\epsilon$, a unitary synthesis algorithm builds a new circuit whose unitary $U_S$ satisfies the inequality $\| U - U_S \| \leq \epsilon$ \cite{younis_qfast_2020}. 
Most recent synthesis tools base their distance metric off the Hilbert-Schmidt inner product as it is computationally inexpensive \cite{davis2020towards, kliuchnikov2014asymptotically, khatri2019quantum}:

\[
    \| U - U_S \|_{HS} = Tr(U^{\dag}U_S)
\]

There are broadly two types of synthesis, top-down and bottom-up. 
Top-down synthesis techniques are rule-based, and aim to break down large unitaries into smaller ones. 
Usually, these algorithms are quick, but the resultant circuit depth grows exponentially, which limits their effectiveness. 
Bottom-up synthesis starts with an empty circuit and  gradually adds gate until a solution is found.
Techniques such as QSearch/LEAP use an A* heuristic search to find an optimal depth approximation for the overall unitary \cite{smith_leap_2021}.

Furthermore, synthesis algorithms such as QSearch also accept a coupling graph as input so that the connectivity between qubits may be specified. 
The coupling graph provided restricts the synthesis algorithm by requiring that it only place multi-qubit gates that obey this connectivity. 
Doing so means that the resulting circuit is fully mapped to the specified coupling graph.
Previous work \cite{smith_leap_2021} has demonstrated that for small circuits, synthesis algorithms are able to map circuits to restrictive qubit topologies using fewer gates than other compiler tools such as \emph{Qiskit} \cite{qiskit} and $t|ket\rangle$ \cite{sivarajah2020t}.
Thus when provided a coupling graph and a small width circuit, synthesis is able to completely remove the need for a mapping algorithm. 
Synthesis algorithms are ultimately limited by the size of the solution search trees, and the size of the unitary matrices for which the Hilbert-Schmidt distance must be calculated.

\subsection{Quantum Circuit Fidelity and Performance}
In the NISQ era, the probability that a quantum circuit is successfully executed depends heavily on the number of multi-qubit gates and the execution time of circuits.
This is because multi-qubit gates have a high probability of introducing noise and because quantum states tend to decohere after a certain amount of time \cite{kandala2019error}.
Producing quantum circuit implementations with both fewer multi-qubit gates and lower depth is therefore desirable, and signals that an implementation has a higher chance of executing properly.
For $N$ partitions and a synthesis threshold $\epsilon=10^{-10}$ per partition, the total circuit approximation error $N \epsilon$ is typically between $10^{-8}$-$10^{-7}$ in Hilbert-Schmidt distance for our selection of benchmarks.
This means although synthesis introduces approximation errors into circuit implementations, the error threshold is sufficiently low that gate and decoherence errors will dominate overall error.
For this reason, we present both CNOT gate counts and circuit depth to evaluate circuit implementations (see Section \ref{section:data}).

Reducing depth is analogous to reducing the run time of a classical program, and thus also acts as a direct measure of circuit performance assuming a quantum machine is capable of executing gates in parallel.
Mitigating other sources of error, such as superconducting crosstalk \cite{Sarovar_2020}, remains an important area of research but is not considered in our evaluation.

\subsection{Mapping Quantum Circuits}
\label{section:mapping_synth_circs}
Quantum circuits are typically designed with the assumption that the underlying hardware that may run the circuit can support interactions between any pair of qubits.
However, due to technological restrictions modern quantum machines typically have very sparse inter-physical qubit connectivity.
Several examples of realistic sparse physical qubit topologies are illustrated in Figure \ref{fig:physical_topologies}.

In order to ensure that a quantum circuit can be run on a quantum computer, a mapping algorithm is used to transform the circuit so that it conforms to some restricted qubit connectivity.
The process of mapping quantum circuits to physical topologies happens in two stages: placement or layout and routing.
During the placement phase, an assignment of logical qubits in the quantum circuit to physical qubits in the qubit topology is created.
In the next phase, the routing phase, quantum SWAP operations are inserted to ensure that all multi-qubit interactions specified in the quantum circuit may take place along edges in the physical topology \cite{li2019tackling, shafaei_optimization_2013}.
Throughout both phases of circuit mapping, the goal is to minimize the number of SWAP operations inserted into the final circuit.
This is because for machines using the \{CNOT, U3\} gate set, a SWAP is implemented using 3 CNOT gates (Figure \ref{fig:swap_gate}).
As quantum circuits become wider and more densely connected, the overhead in CNOT count due to routing SWAP gates quickly rises.
However, minimizing the number of SWAP operations is known to be NP-Hard \cite{siraichi2018collange}, so most algorithms use heuristic-based approaches that are sub-optimal, especially for wider circuits.

\subsection{Quantum Circuit Partitioning}
\label{section:quantum_circ_partitioning}
\begin{figure}
    \centering
    \includegraphics[width=8cm]{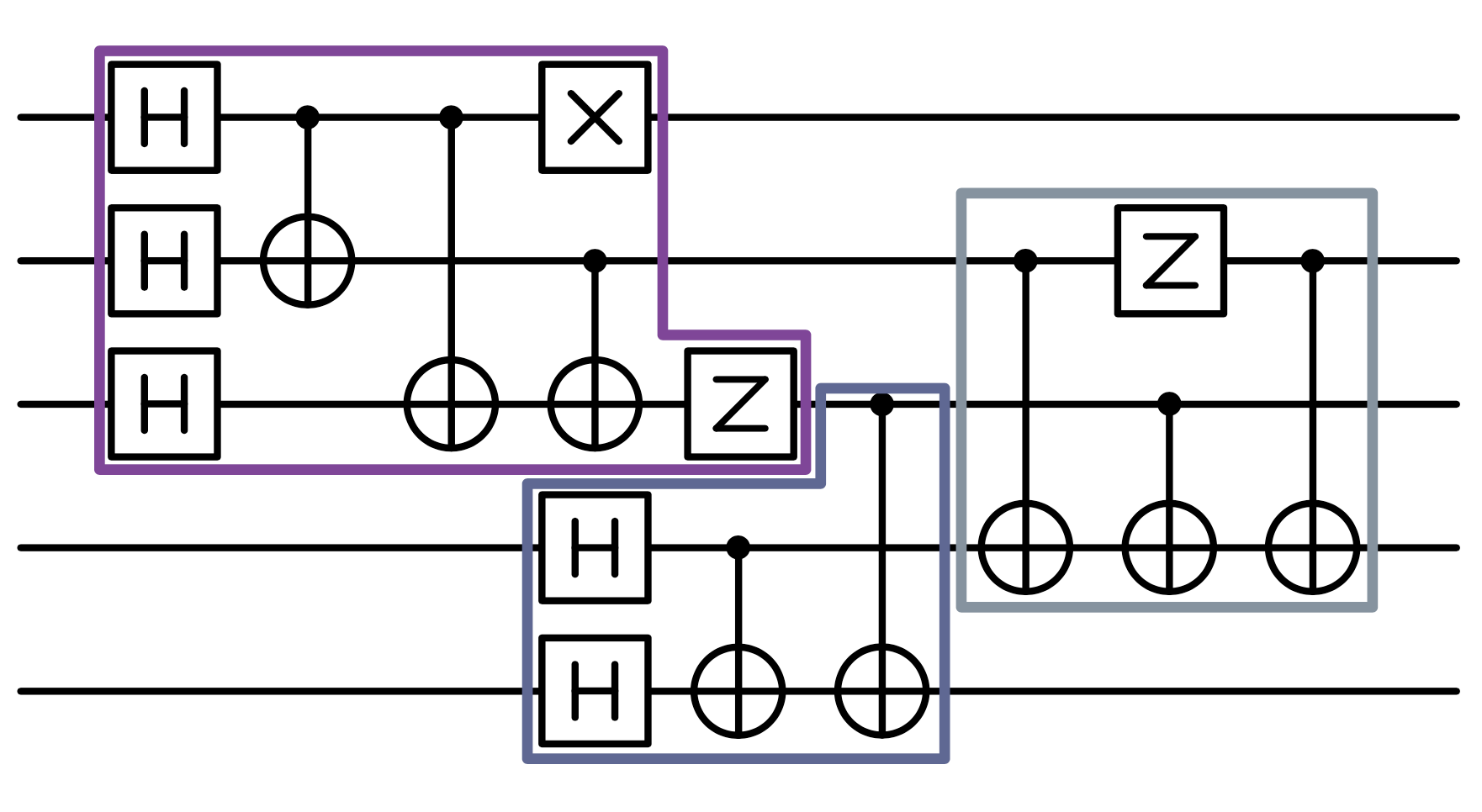}
    \caption{
        Example quantum circuit with 5 qubits. 
        Each qubit is represented by a horizontal line, and the timing of the circuit is described by the x-axis.
        The circuit has been partitioned into 3 subcircuits.
    }
    \label{fig:q_circuit}
\end{figure}

Given a quantum circuit and a positive integer $k$ (called the partition width or block size), a partitioning algorithm divides the circuit into subcircuits (also called blocks or partitions) of width at most $k$. Valid partitions consist only of gates that act on those specific qubits in the partition. 
As soon as a gate crosses a partition boundary, the partition must be cut vertically along the violating wire. 
A new partition can then be created, or more gates can be added so long as they act only on qubits still within the partition. 
An example circuit with width 3 partitions is shown in Figure \ref{fig:q_circuit}.

For the purposes of unitary synthesis, the goal of a partitioning algorithm is to form as few partitions as possible, and thus ensure that each of these partitions is as large as possible.
Our experiments show that these large partitions tend to see a larger reduction in CNOTs as compared to smaller partitions.
This effect is illustrated in Figure \ref{fig:partition_summary}.

\subsection{Post-Mapping vs. Logical Circuit Synthesis}
\begin{figure}
    \centering
    \begin{subfigure}[t]{0.45\textwidth}
        \centering
        \includegraphics[width=6cm]{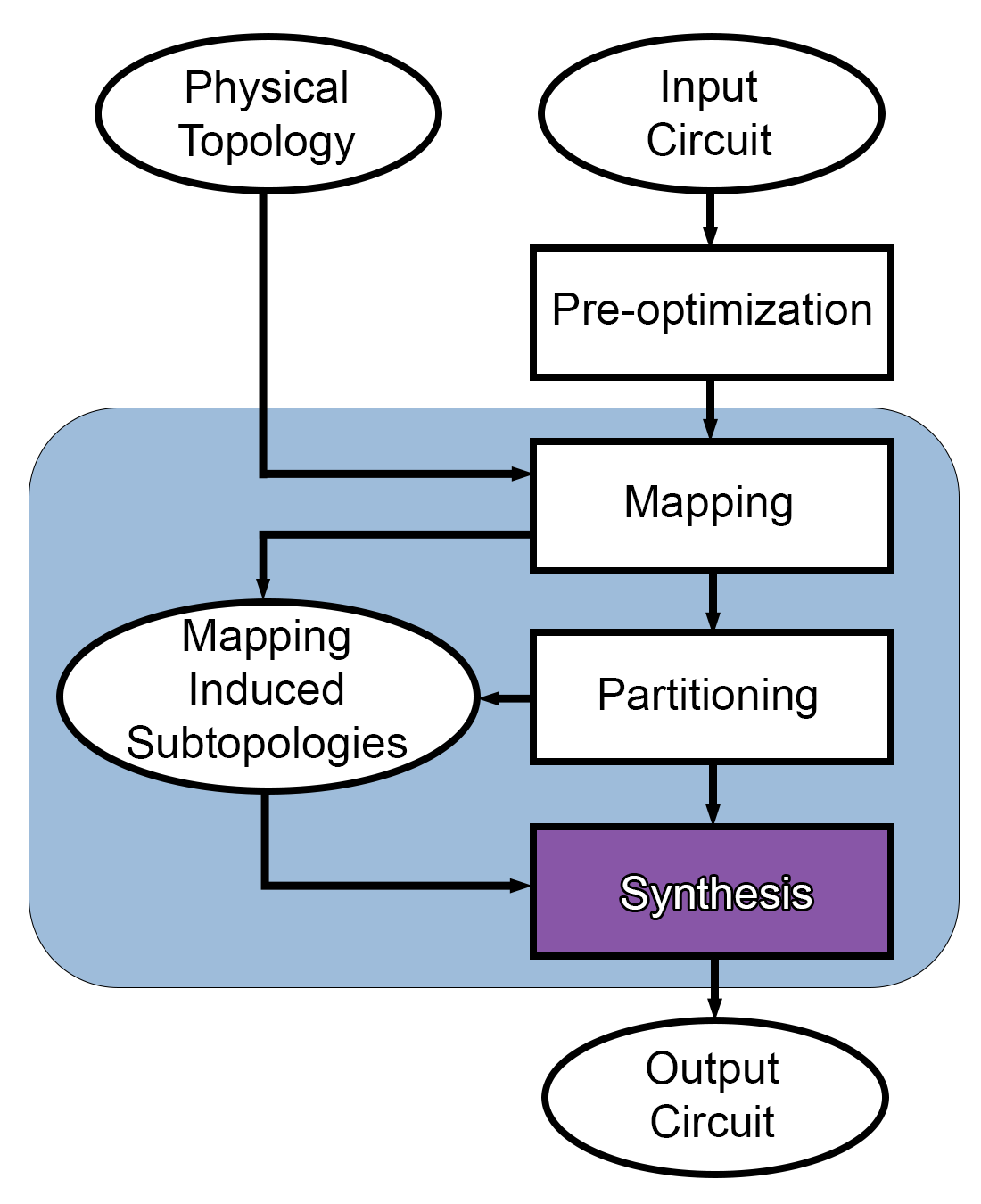}
        \caption{Post-mapping synthesis optimization.}
        \label{fig:qgo_flow}
    \end{subfigure}
    \begin{subfigure}[t]{0.45\textwidth}
        \centering
        \includegraphics[width=6cm]{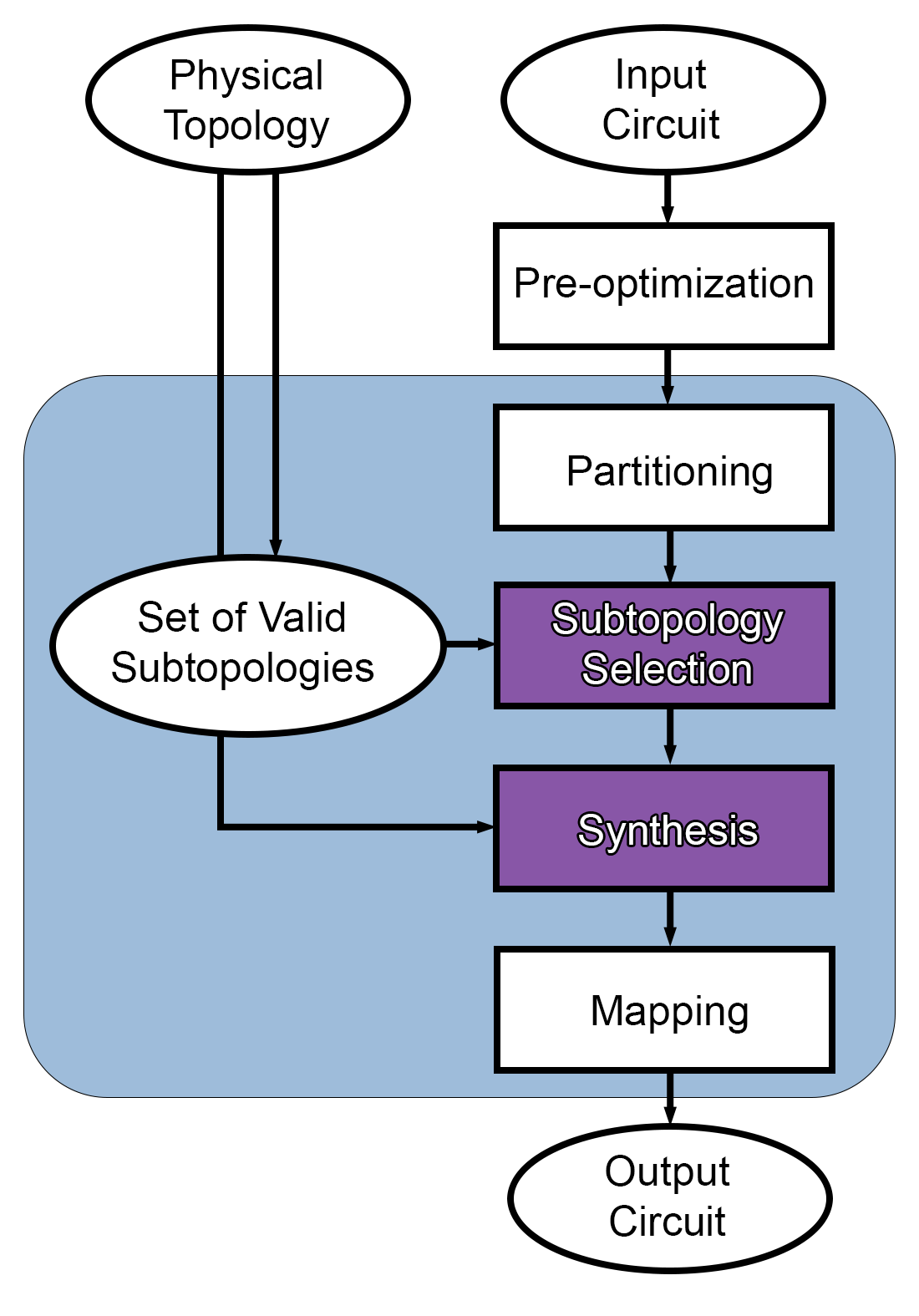}
        \caption{Pre-mapping or logical circuit synthesis optimization.}
        \label{fig:topas_flow}
    \end{subfigure}
    \caption{Program flow diagrams for wide circuit optimization tools using unitary synthesis post- and pre-mapping.}
\end{figure}

Previous tools to apply unitary synthesis optimization to wide quantum circuits, such as QGo \cite{qgo}, follow a \emph{post-mapping} synthesis flow as illustrated in Figure \ref{fig:qgo_flow}. 
After preliminary circuit optimizations, the logical circuit is mapped to a specified physical qubit topology. 
This mapped quantum circuit is then partitioned into subcircuits, each of which are independently synthesized. 
The synthesized subcircuits are then reassembled into the complete optimized and mapped circuit.

Although post-mapping synthesis schemes are able to reduce depth and multi-qubit gate count, there are several pitfalls that limit their effectiveness.
First, because mapped circuits typically contain more gates than their unmapped counterparts, partitioning algorithms tend to form far more partitions on mapped circuits.
Because each of these partitions must be individually synthesized, this approach is typically more time consuming than partitioning and synthesizing the logical, unmapped, circuit.
Second, these methods are very sensitive to the quality of the mapping algorithm used, as they can only reduce inefficiencies introduced into mapped circuits.
Doing the same on the logical circuit instead allows for mapping to be done on an (often much) shorter depth and gate count approximate circuit. 
Logically synthesized circuits typically have fewer gates, which means fewer SWAP gates are typically needed during routing.

These points motivate the reasoning behind adopting a pre-mapping, or logical circuit synthesis flow.
The process of logical circuit synthesis for wide quantum circuits is illustrated in Figure \ref{fig:topas_flow}. 
After some initial quick circuit optimizations, the logical circuit is partitioned into synthesizable subcircuits.
Each of these subcircuits is paired with a synthesis subtopology to which the subcircuits are mapped.
After each subcircuit is synthesized, the circuit is reconstructed and finally fully mapped to the specified physical qubit topology.

\subsection{Synthesis Subtopology Selection}
\label{section:synth_sub_selection}
Each partitioned subcircuit must be assigned a graph $G_S = (V, E_S)$ that specifies the connectivity between qubits in the partition.
The synthesis algorithm chooses multi-qubit gates that correspond to edges in the set $E_S$, mapping the subcircuit to this \emph{synthesis subtopology} $G_S$. 
The synthesis subtopology selection process in logical circuit synthesis tools diverges from that in post-mapping synthesis tools such as QGo. 
Since post-mapping tools assume circuits are already routed, the interactions between the qubits in a partition are guaranteed to fit the physical topology. 
Therefore, we can choose $G_S$ to be the subgraph induced by the physical qubits in a partition.

Logical synthesis affords more choice here as the partitioned circuit is not yet routed to obey the restrictions of the physical hardware.
Each synthesis subtopology is a graph of order equal to the number of qubits in that partition.
The only other requirement for synthesis subtopologies is that they are connected.
This is because if two qubits interact in the partitioned circuit, but there is no path between them in the synthesis subtopology, the synthesis algorithm will be unable to find a circuit whose unitary approximates the original subcircuit's unitary.
As synthesis algorithms are mostly limited to operating in the 3-5 qubit partition width ranges, the number of possible synthesis subtopologies is relatively limited.

Our experiments show that the choice of synthesis subtopology can have drastic effects on the number of multi-qubit gates in the optimized subcircuit, as well as on the performance of the routing algorithm during the final mapping process.
The goal of the synthesis subtopology selection process is thus to choose a graph $G_S$ for each partition such that the synthesis algorithm is able to produce low gate count subcircuits and the mapping algorithm is able to more efficiently place and route the subcircuit. 
The structure of the subtopology $G_S$ also heavily impacts the depth of synthesized subcircuits.
Depth is heavily dependent on the degrees of vertices in the subtopology $G_S$.
Subtopology selection thus also must carefully consider the connectivity within a partition to ensure that the depth of the optimized circuits is minimized.
Section \ref{section:top_aware_subtopology_selection} further details the choices made for the subtopology selection algorithm for the TopAS tool.

\section{Topology Aware Synthesis}
\label{section:topas}

We present TopAS, a physical qubit topology aware synthesis tool.
By partitioning and synthesizing subcircuits from the logical circuit, the number of multi-qubit operations in a quantum circuit is reduced before mapping begins.
The choice to synthesize logical quantum circuits enables TopAS to produce mapped circuits with fewer total CNOT gates than many other optimization tools.
TopAS selects synthesis subtopologies for partitioned subcircuits that are sparse and easily embedded within the underlying physical qubit topology on which the circuit will be run. 
Synthesis subtopology selection is done in such a way that reduces the number of operations needed for both computation and mapping.

The TopAS tool uses a \emph{scan} partitioning strategy introduced by the authors of QGo \cite{qgo}. 
The synthesis algorithm used is the QSearch/LEAP algorithm \cite{smith_leap_2021}.
Working versions of the partitioning and synthesis algorithms are provided by the \emph{BQSKit} tool \cite{bqskit}. 
A flowchart of the TopAS tool's execution is displayed in Figure \ref{fig:topas_flow}.

\subsection{Logical Quantum Circuit Partitioning}
\label{section:scan_partitioner}
\begin{figure}
    \centering
    \includegraphics[width=0.48\textwidth]{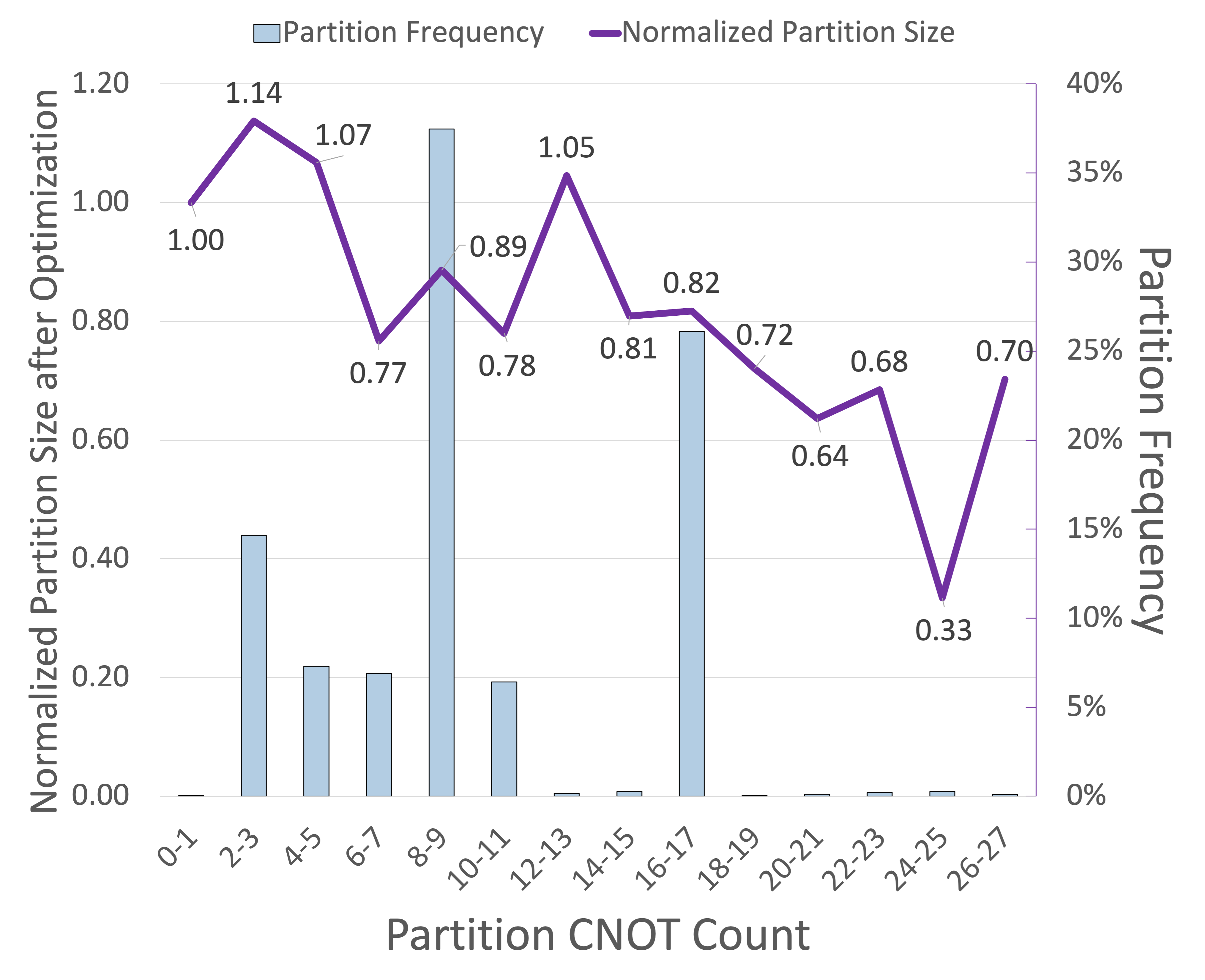}
    \caption{
        Partition frequency and mean normalized partition size after optimization as a function of CNOT count.
        Normalized partition size is calculated using the number of CNOTs in the synthesized subcircuit divided by CNOTs in the original partition.
        A value less than 1 indicates that synthesis reduces those partitions.
        The partitioner predominately forms subcircuits with 6-11 CNOTs, but is also able to form many subcircuits with 16-17 CNOTs.
    }
    \label{fig:partition_summary}
\end{figure}

The partitioning algorithm views the input circuit as a two dimensional grid, where qubits are represented by rows, and time steps in the circuit are represented by columns.
Each element of the grid either contains a quantum operation or is empty.
The width of this grid is $n$, the number of qubits in the circuit.
The length is $T$, the depth or critical path length of the circuit.
The \emph{scan} partitioner algorithm sequentially \emph{scans} through all as yet unpartitioned gates in a circuit, examines each possible grouping of $k$ or fewer qubits, and picks the one that lends the largest partition.
The \emph{scan} partitioner tends to form few partitions, and often forms partitions with a high average number of multi-qubit gates.
As mentioned in Section \ref{section:quantum_circ_partitioning} and illustrated in Figure \ref{fig:partition_summary}, subcircuits with a large number of multi-qubit gates tend to yield the most reduction by synthesis.
The distribution of partitions produced by the \emph{scan} partitioning algorithm for a subset of benchmarks is also shown in Figure \ref{fig:partition_summary}.

Note that partitions consist only of subcircuits with interacting qubits.
A circuit logical connectivity graph is used to enforce this policy.
This graph has a vertex for each qubit in the circuit.
The presence of an edge $(u,v)$ indicates that there is a multi-qubit gate between qubits $u$ and $v$ at some point in the circuit.
Given a partition width of $k$, candidate partitions consist of all connected subgraphs in this circuit logical connectivity graph of order at most $k$.
In the worst case, a circuit where all qubits interact with each other, there are $O{n \choose k}$ candidate partitions.

\subsection{Synthesizing Logical Quantum Circuits}
\label{section:synth_logical_quantum_circs}
As logical circuits do not contain gates needed to conform to some restrictive physical topology, they tend to be shorter than their mapped counterparts.
Shorter circuits tend to yield fewer partitioned subcircuits, as there are fewer total gates to partition.
Partitioning circuits before mapping thus conserves the number of subcircuits produced by the partitioning algorithm.
As fewer subcircuits must be independently synthesized in this case, less time is often required to optimize circuits that are partitioned before mapping.
Circuits with fewer partitions also tend to have more accurate output circuits.
Each partitioned circuit is synthesized independently, so each synthesis procedure will produce a mapped circuit whose unitary representation will be within some Hilbert-Schmidt norm distance $\epsilon$ away from the target unitary.
The authors of \cite{quest} showed that the total error in a circuit that consists of synthesized subcircuits is bounded by the sum of each subcircuit's distance.
If a circuit is partitioned into $N$ subcircuits, because each partition is synthesized to within a distance $\epsilon$, the total distance of the partitioned and synthesized circuit is bounded by $N \epsilon$.
Decreasing $N$, as is done in the logical partitioning case, thus improves the accuracy of the synthesized circuit.
Results comparing the upper bound on total circuit errors for several synthesized benchmarks are shown in Table \ref{table:error_and_avg_cnots}.

\subsection{Topology Aware Subtopology Selection}
\label{section:top_aware_subtopology_selection}
The main contribution of this work is a subtopology selection strategy and compilation workflow built around it. 
It allows for partitioned subcircuits to be preconditioned in such a way that balances the opposing demands of synthesis and mapping.
As discussed in Section \ref{section:synth_sub_selection}, the goal of the subtopology selection process is to assign a graph $G_S$ to each partitioned subcircuit.
The vertices in $G_S$ represent qubits in the subcircuit, while edges describe the allowed interactions between qubits in the output synthesized subcircuit.
\begin{figure}
    \centering
    \includegraphics[width=.48\textwidth]{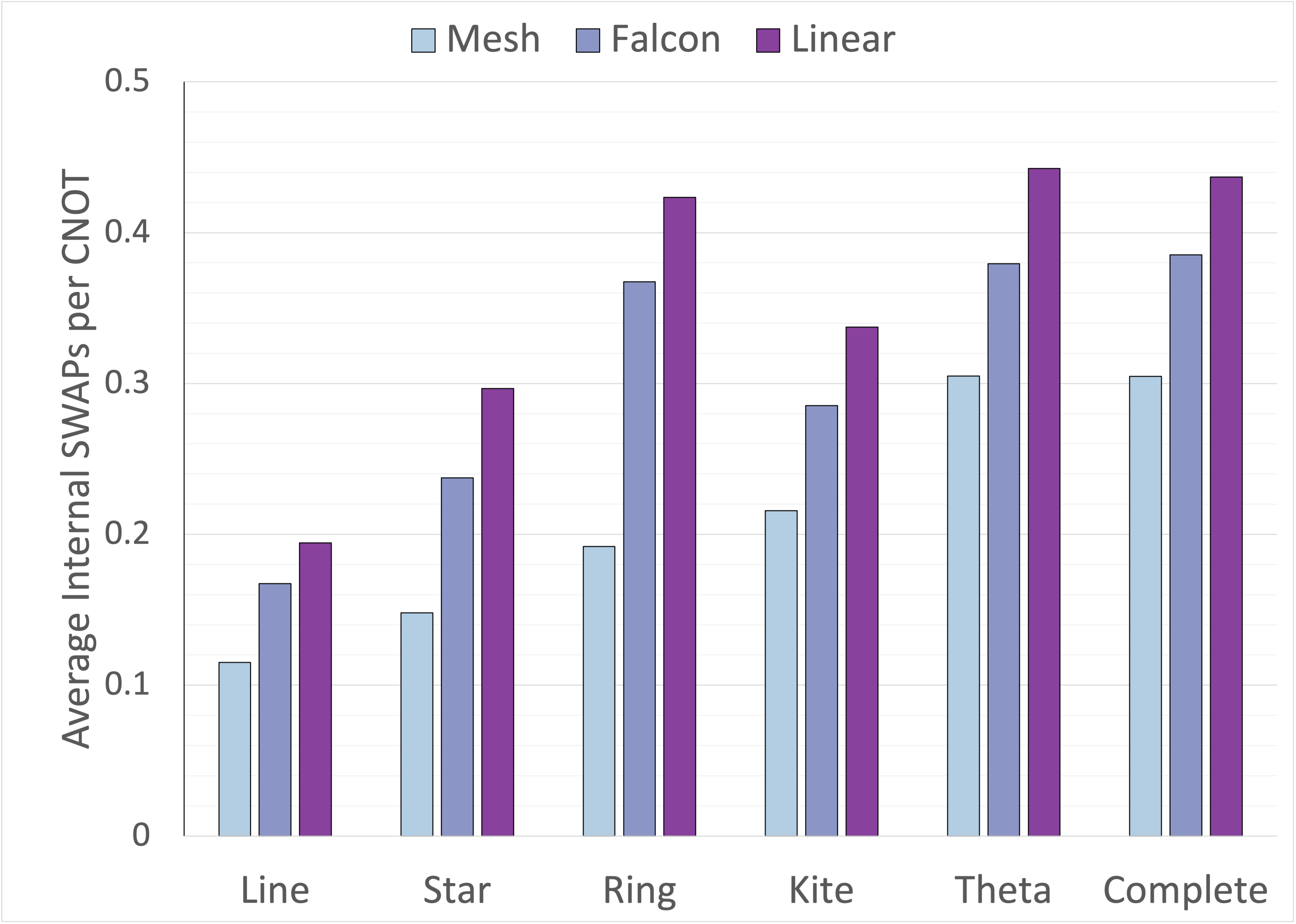}
    \caption{
        The number of SWAP operations within partitions synthesized to different synthesis subtopologies for the mesh, falcon, and linear physical topologies.
        Simpler subtopologies result in fewer SWAP, but may produce circuits with more CNOTs.
    }
    \label{fig:internal_swaps}
\end{figure}
\begin{figure}
            \subcaptionbox{Line \label{line}}[.2\textwidth]{\includegraphics[width=.1\textwidth]{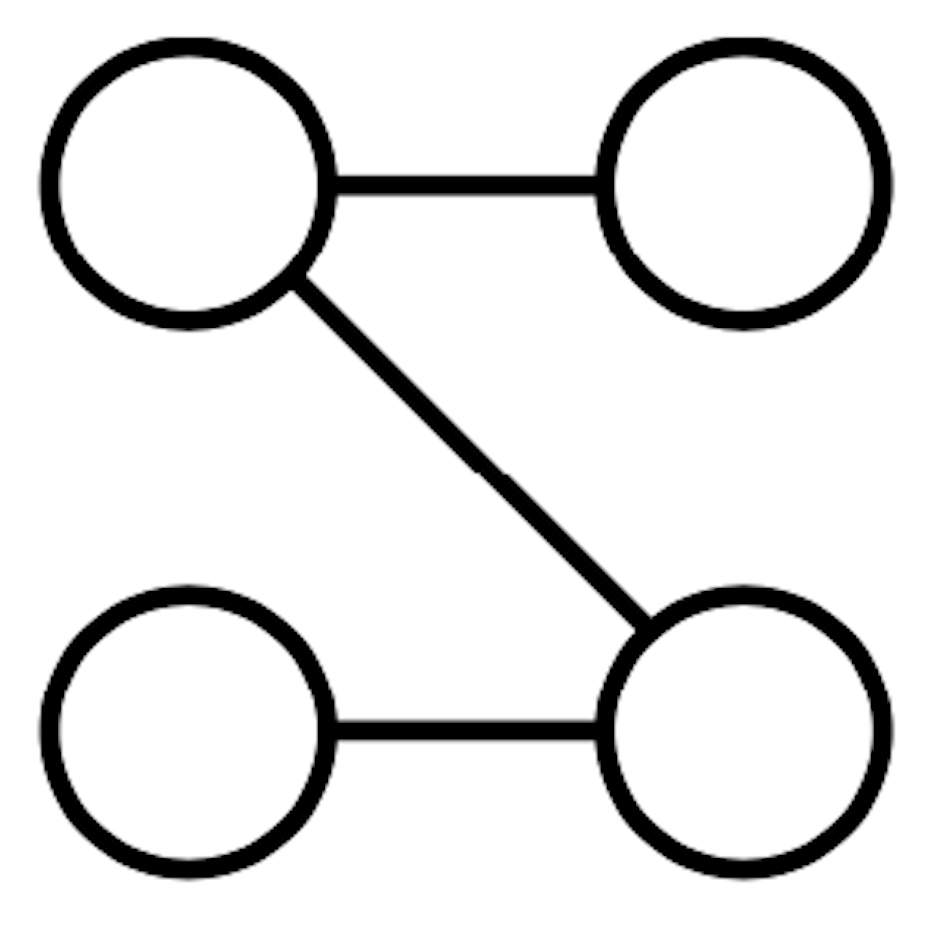}}
            \subcaptionbox{Star \label{star}}[.3\textwidth]{\includegraphics[width=.1\textwidth]{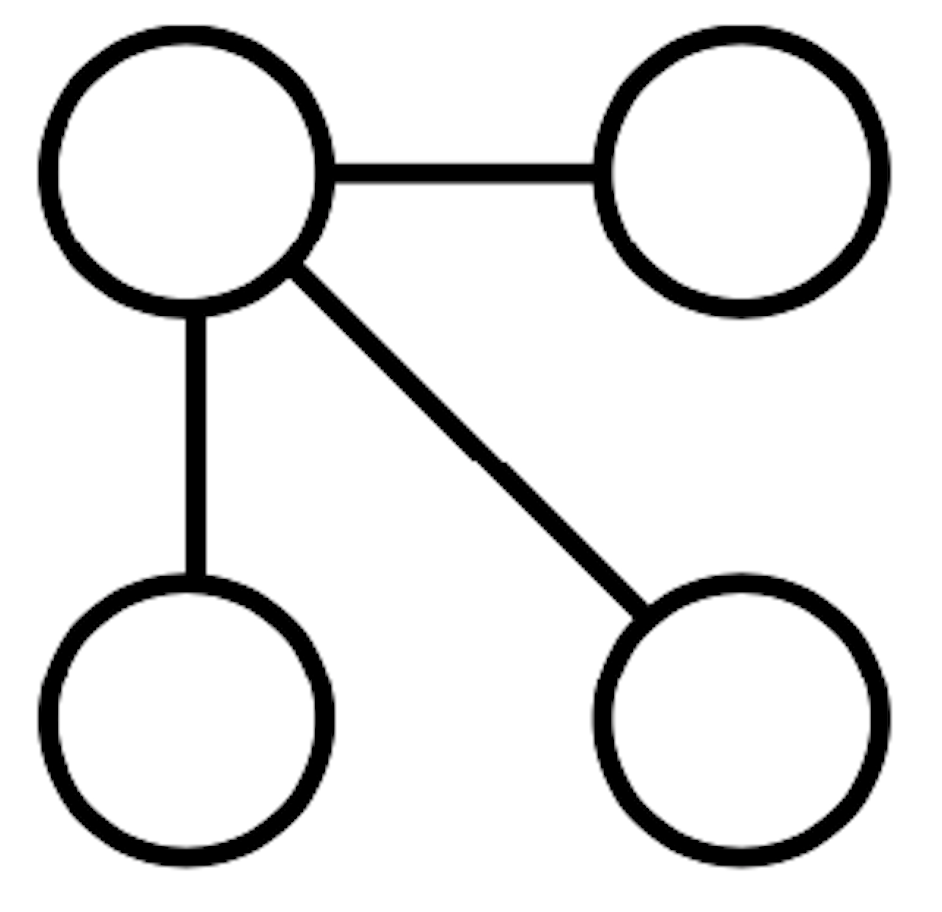}}
            \subcaptionbox{Ring \label{ring}}[.2\textwidth]{\includegraphics[width=.1\textwidth]{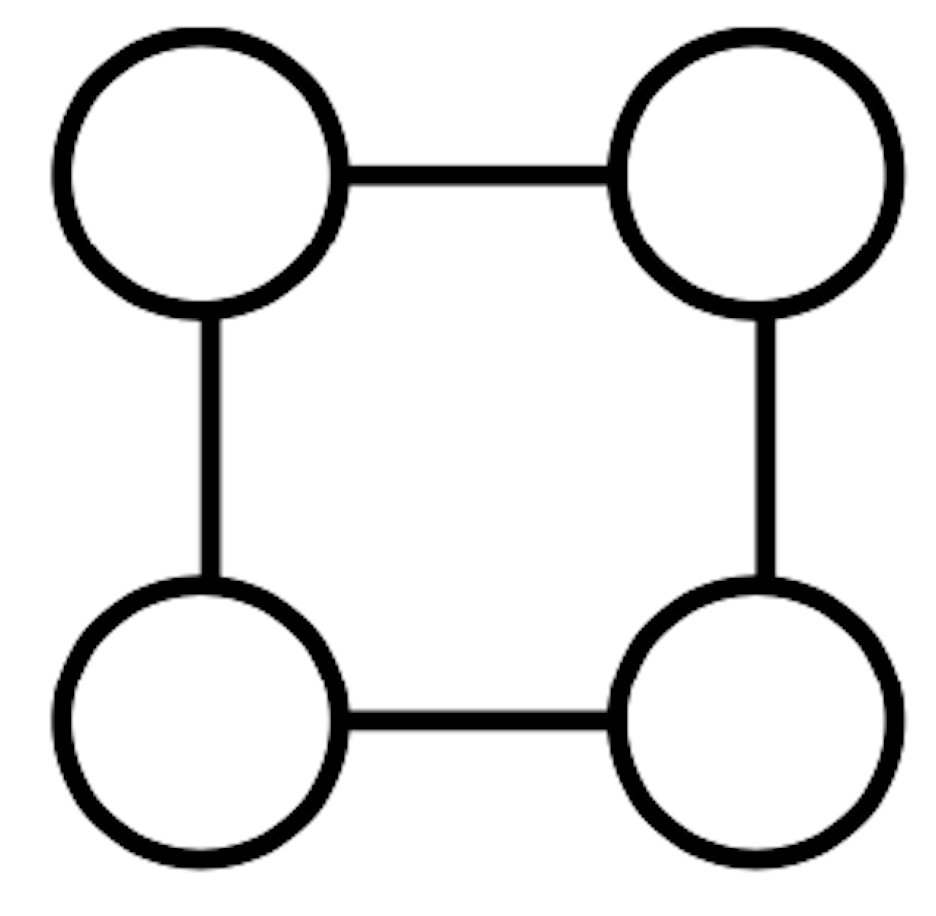}}
            \subcaptionbox{Kite \label{kite}}[.3\textwidth]{\includegraphics[width=.1\textwidth]{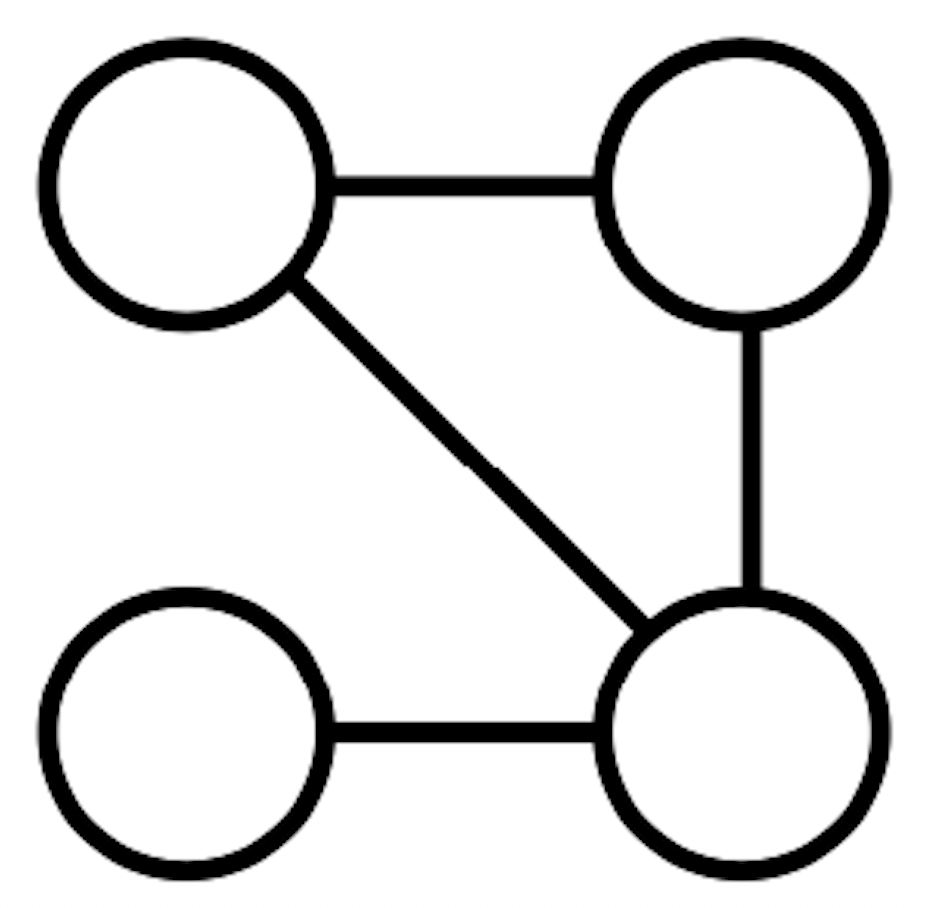}}
          \subcaptionbox{Theta \label{theta}}[.2\textwidth]{\includegraphics[width=.1\textwidth]{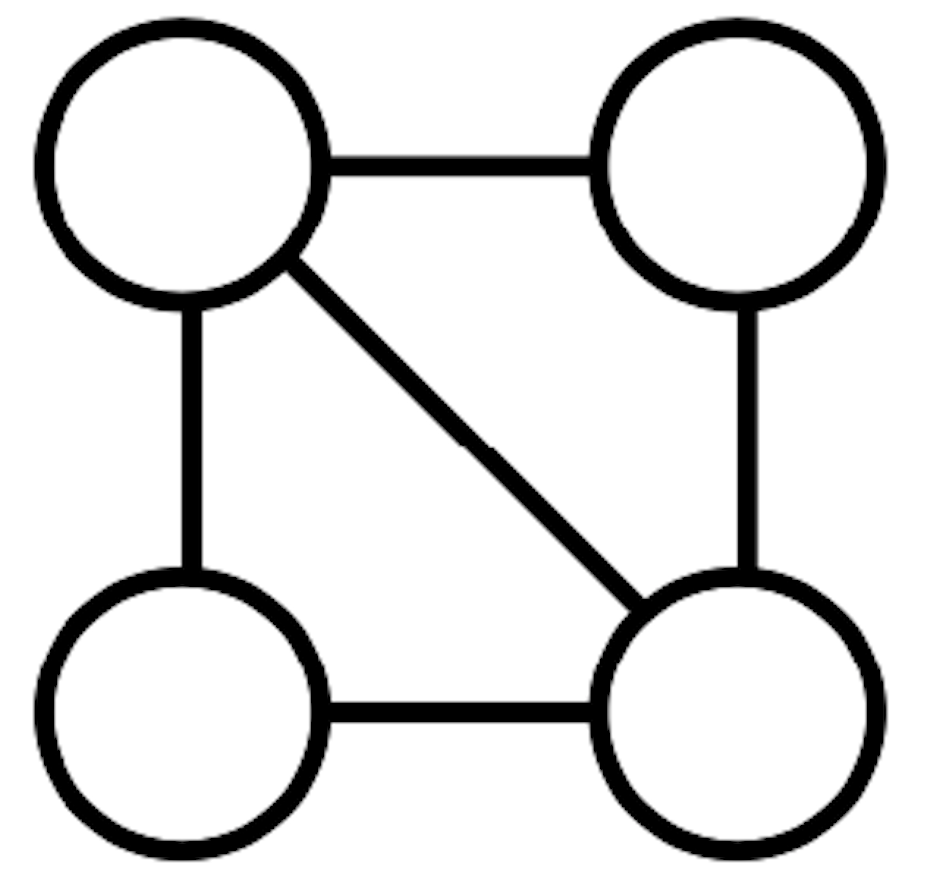}}
    \subcaptionbox{Complete \label{complete}}[.3\textwidth]{\includegraphics[width=.1\textwidth]{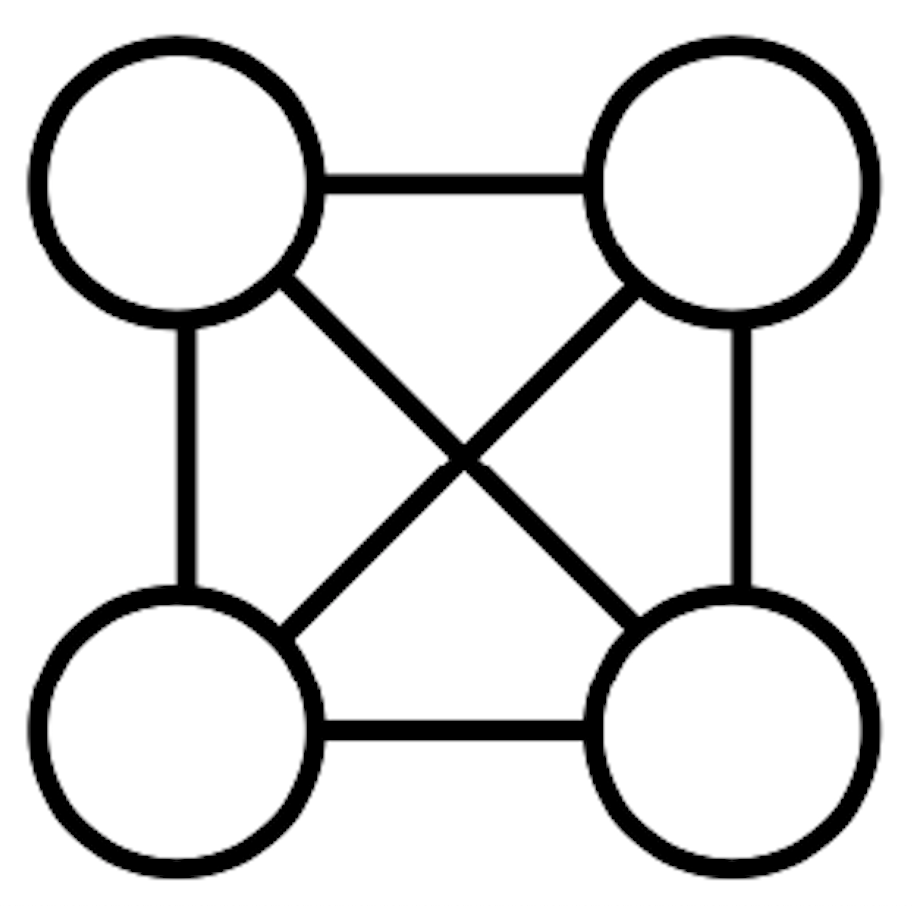}}
    \caption{
        Possible order 4 synthesis subtopologies. The graphs a, b, and c are embedded within the mesh physical topology. Graphs d, e, and f are not. Only a and b are embedded in the falcon topology, while a is the only embedded subgraph of the linear topology.
    }
    \label{fig:subtopologies}
\end{figure}
\begin{figure}
    \centering
    \includegraphics[width=.2\textwidth]{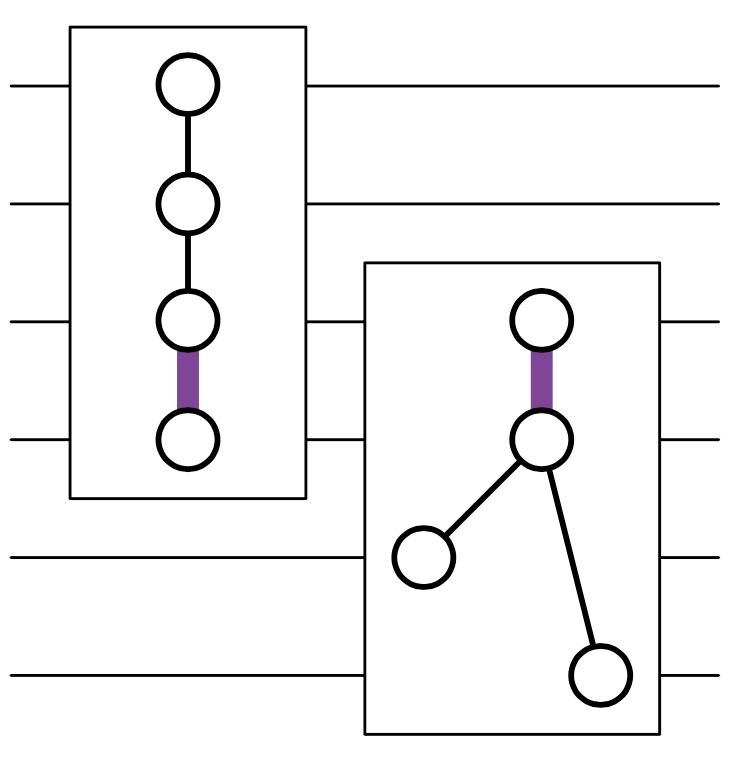}
    \caption{Neighbor aware subtopology selection encourages the reuse of edges between sequentially executing partitions.}
    \label{fig:neighbor_aware}
\end{figure}

%
Ideally, a quantum circuit is synthesized so that it uses only edges present in the physical topology for multi-qubit interactions.
Before mapping, the placement and thus allowed interactions between qubits is not known.
However, synthesizing densely connected subcircuits to sparse graph structures can decrease the need for SWAP operations during the mapping phase.
Limiting the set of synthesis subtopologies and carefully selecting sparse subtopologies for each subcircuit allows us to balance the opposing demands of the routing and synthesis algorithms, leading to fewer total CNOT gates.
Figure \ref{fig:internal_swaps} shows the average number of internal SWAP gates per CNOT for a variety of synthesis subtopologies targeting three physical topologies.
Internal SWAP gates are routing operations that are inserted between the first and last gate of a partition.
If each subcircuit is synthesized to an embedded subgraph of the physical topology and is executed atomically, there is a solution to the mapping problem that requires zero SWAP operations on the subcircuit's qubits during its execution.
However, because routing often disrupts the execution of subcircuits, it is unlikely for there to be exactly zero internal SWAP operations even for subtopologies that are embedded within the physical topology.
The number of internal SWAP gates observed for a partition is normalized by the number of CNOTs in the partition.
This allows us to more effectively compare subcircuits of various sizes.
Typically, sparser subtopologies require more gates for computation than denser subtopologies, but they require fewer SWAP gates during mapping.

The TopAS tool restricts the set of graphs from which to choose synthesis subtopologies to the set of connected graphs that are embedded in the target physical topology.
TopAS uses a partition width of 4 qubits.
The order 4 candidate subtopologies are illustrated in Figure \ref{fig:subtopologies}.

QSearch, the synthesis algorithm at the core of the TopAS optimization flow, aims to synthesize circuits using as few multi-qubit gates as possible.
Often times, circuit depth is minimized as a consequence of this goal  but is not the primary metric of success.
Subtopology selection is the only variable that TopAS uses to reduce circuit depth directly.
For certain subtopologies choices, circuits produced tend to contain many gates that must be executed sequentially.
The most significant contributor to this property is the possible graph matchings present in a given subtopology.
In topologies such as the star (Figure \ref{fig:subtopologies}b), any choice of a single edge is a maximum matching.
This means that for the star subtopology, no two CNOTs can be executed in parallel.
Other subtopologies like the line and ring (Figure \ref{fig:subtopologies}a and \ref{fig:subtopologies}c) have larger maximum matchings, and thus allow more parallelism within synthesized subcircuits.
Although they typically require more routing gates, subtopologies containing more edges tend to allow for lower depth circuits.

The qubit interactions within a partitioned subcircuit can be described using a weighted undirected graph $G_L = (V, E_L)$.
For each multi-qubit interaction that occurs between qubits $k_1, k_2 \in [K]$ in the subcircuit, there is an edge $(k_1, k_2, w) \in E_L$, where the weight $w$ corresponds to the number of times that interaction occurs.
Synthesis algorithms such as QSearch allow for the connectivity between qubits to be specified by an unweighted undirected graph $G_S = (V,E_S)$.
For a maximum partition size of $k$, $V=[k]$ and $E_S \subseteq V \times V$.

Our experiments indicate that synthesis is best able to reduce the multi-qubit gate count of subcircuits when the logical connectivities and synthesis subtopologies match.
A kernel function \cite{hofmann_kernel_2008} is used to quantify the similarity between the graphs.
The scoring function $K:G_L \times G_S \rightarrow [0, 1]$ examines each edge in the synthesis subtopology $G_S$ and logical connectivity $G_L$.
From the edges, two vectors $v_L, v_P \in R^{k(k-1)/2}$ are constructed.
The vector $v_P$ is simply an indicator vector, with a $1$ at each element that corresponds to a present edge in $G_S$.
Element $i$ of $v_L$ contains the weight $w$ associated with edge $i$ in $G_L$.
The normalized inner product 
$$
    \text{similarity}(v_P, v_L) = \frac{v_P^T v_L}{\sum\limits_{i} v_{L}(i)} 
$$ 
is then returned to quantify the similarity between the graphs.
For each partition, all $k!$ permutations of qubit labels are evaluated for each of the candidate synthesis subtopologies.
The permuted subtopology with the highest kernel function score is considered the best candidate subtopology for the purposes of producing the smallest output subcircuit.
The effectiveness of the similarity kernel function was evaluated by synthesizing partitions of width 4 to all order 4 subgraphs embedded within the mesh physical topology (Figures \ref{fig:subtopologies}a, \ref{fig:subtopologies}b, and \ref{fig:subtopologies}c). 
In this scenario, the similarity function was able to identify the subtopology that would produce the synthesized subcircuits with the fewest CNOT gates 89\% of the time.

If multiple subtopologies have the same similarity score, that with the fewest edges is preferred.
In order favor sparse subtopologies, each similarity is multiplied by a bias factor.
The biases that produced the best results for the mesh topology were $1.0$ (line), $1.0$ (star), and $0.8$ (ring).

When selecting subtopologies, TopAS also considers the impact of subtopology choice for the immediately preceding and succeeding partitions.
Although partitions are synthesized separately, they are not executed in isolation.
If a partition assumes a physical edge exists between two qubits, and the partition immediately following also assumes this physical edge exists, fewer total SWAPs may be needed between the execution of these two partitions.
The TopAS tool therefore checks each subcircuit's neighboring partitions for qubits that are shared.
If interactions from the neighboring partitions occur between shared qubits, they are added as edges to the current partition's logical connectivity graph $G_L$.
Interactions from neighboring partitions carry the same weight as interactions within a partition, even if the qubits in the neighbor partitions do not interact directly within the partition itself.
This policy was chosen as larger amounts of edge sharing between partitions greatly reduced the final CNOT count of optimized circuits.
TopAS uses this \emph{neighbor aware} version of the logical connectivity graph in the computation of the similarity function.
An example of two sequentially executing partitions with subtopologies that have been selected to reuse an edge is shown in Figure \ref{fig:neighbor_aware}.

\begin{algorithm}
    \caption{Subtopology Selection}
    \textbf{Input}: graph $G_L$, set of graphs $G_{neighbors}$, $bias$ function \\
    \textbf{Output}: graph $G_S$
    
    \begin{algorithmic}
        \FORALL {graphs $G_n$ in $G_{neighbors}$}
            \IF {$shared\_edge$ in $G_n$ and in $G_L$}
                \STATE {$G_{overlap} \leftarrow shared\_edge$}
            \ENDIF
        \ENDFOR
        \STATE {$v_L \gets$ edge weight vector for $G_{overlap} \cup G_L$} 
        \STATE {$candidates \gets$ order $\leq 4$ subgraphs in physical topology}
        \FORALL{graphs $G_P$ in $candidates$}
            \STATE {$v_P \gets$ indicator vector for $G_P$}
            \STATE {$score = $similarity$(v_P, v_L)$}
            \IF {$bias(G_P) \times score > high\_score$}
                \STATE {$high\_score \gets bias(G_P) \times score$}
                \STATE {$G_S \gets G_P$}
            \ENDIF
        \ENDFOR
        \RETURN $G_S$
    \end{algorithmic}
\end{algorithm}

\subsection{Partition Replacement and Mapping}
As indicated in Figure \ref{fig:partition_summary}, it is sometimes the case that synthesized subcircuits grow instead of reduce in size.
When this happens, TopAS considers replacing the synthesized subcircuit with the original subcircuit.
Always choosing the logical subcircuit when it has fewer CNOTs can be problematic in the final routing pass, as the original subcircuit may be more difficult to route than the synthesized version.
Thus, TopAS considers differences in both the number of CNOTs and logical connectivities of the two subcircuits.
TopAS adopts a policy of replacing the synthesized subcircuit with the original if it shows more than 30\% (empirically determined best threshold) fewer CNOTs or has a more routable logical connectivity.
The routability of a subcircuit's logical connectivity is determined by the average number of of internal SWAP operations per CNOT for a given physical topology (see Figure \ref{fig:internal_swaps}).

After subtopology selection and synthesis take place, TopAS maps the synthesized quantum circuit to a physical topology.
For placement and routing, the SABRE Layout and SABRE Swap \cite{li2019tackling} mapping algorithms are used.

\section{Results}
\label{section:data}
\begin{figure}[ht]
    \small
    \centering
    \begin{tabular}{|c|c|c|c|c|}
    \hline
                    &         &          & Total             & CNOTs per     \\ 
                    & Mapping & Topology & Error             & Partition     \\ \hline \hline
                    
                    & QGo     &  Falcon  & 1.38E-08          & 15.3          \\ \cline{2-5}
        mult 16     & QGo     &  Mesh    & 1.51E-08          & 12.9          \\ \cline{2-5}
                    & TopAS   &     -    & \textbf{7.20E-09} & \textbf{15.4} \\ \hline \hline
                    & QGo     &  Falcon  & 1.17E-07          & 12.5          \\ \cline{2-5}
        mult\_32    & QGo     &  Mesh    & 1.15E-07          & 11.7          \\ \cline{2-5}
                    & TopAS   &     -    & \textbf{4.96E-08} & \textbf{15.1} \\ \hline \hline
                    & QGo     &  Falcon  & 5.83E-08          & 7.3           \\ \cline{2-5}
        qft\_64     & QGo     &  Mesh    & 5.18E-08          & 7.8           \\ \cline{2-5}
                    & TopAS   &    -     & \textbf{2.32E-08} & \textbf{8.1}  \\ \hline \hline
                    & QGo     &  Falcon  & 8.09E-08          & \textbf{8.1}  \\ \cline{2-5}
        qft\_100    & QGo     &  Mesh    & 8.72E-08          & 7.4           \\ \cline{2-5}
                    & TopAS   &    -     & \textbf{3.85E-08} & 8.0           \\ \hline \hline
                    & QGo     &  Falcon  & 9.09E-08          & 6.5           \\ \cline{2-5}
        add\_65     & QGo     &  Mesh    & 7.02E-08          & \textbf{7.3}  \\ \cline{2-5}
                    & TopAS   &          & \textbf{3.53E-08} & 7.0           \\ \hline \hline 
                    & QGo     &  Falcon  & 1.26E-07          & \textbf{11.8} \\ \cline{2-5}
        add\_101    & QGo     &  Mesh    & 1.23E-07          & 9.4           \\ \cline{2-5}
                    & TopAS   &    -     & \textbf{6.18E-08} & 7.0           \\ \hline \hline 
                    & QGo     &  Falcon  & 1.42E-07          & 7.61          \\ \cline{2-5}
        tfim\_40    & QGo     &  Mesh    & 1.57E-07          & 6.3           \\ \cline{2-5}
                    & TopAS   &    -     & \textbf{1.05E-07} & \textbf{8.1}  \\ \hline \hline
                    & QGo     &  Falcon  & 5.17E-07          & \textbf{8.4}  \\ \cline{2-5}
        tfim\_100   & QGo     &  Mesh    & 4.71E-07          & 7.7           \\ \cline{2-5}
                    & TopAS   &     -    & \textbf{2.70E-07} & 8.0           \\ \hline \hline 
                    & QGo     &  Falcon  & 1.83E-07          & \textbf{8.0}  \\ \cline{2-5} 
        hubbard\_18 & QGo     &  Mesh    & 1.60E-07          & 6.5           \\ \cline{2-5}
                    & TopAS   &          & \textbf{9.56E-08} & 3.7           \\ \hline \hline
                    & QGo     &  Falcon  & 3.86E-07          & 10.3          \\ \cline{2-5}
        shor\_26    & QGo     &  Mesh    & 3.32E-07          & 11.6          \\ \cline{2-5}
                    & TopAS   &     -    & \textbf{1.52E-07} & \textbf{13.9} \\ \hline
    \end{tabular}
    \captionof{table}{
        Upper bound of total circuit error (sum of Hilbert-Schmidt distances per partition) and average CNOTs per partition for each benchmark.
        A synthesis threshold of $\epsilon = 10^{-10}$ was used for both QGo and TopAS.
    }
    \label{table:error_and_avg_cnots}
\end{figure}

The performance of TopAS is compared to other optimization and mapping tools by measuring the depth and total CNOT operation counts of optimized and mapped circuits.
Here, we compare TopAS (as described in Section \ref{section:topas}) to the \emph{Qiskit}, $t|ket\rangle$, and QGo tools.
In all cases, circuits are optimized using $t|ket\rangle$'s full peephole optimization.
Afterwards, either \emph{Qiskit} or $t|ket\rangle$ are used to map the quantum circuit to the specified physical topology.
The QGo tool was used to optimize either the \emph{Qiskit} or $t|ket\rangle$ mapped circuits, whichever contained fewer CNOT operations.
The synthesis tool used is the QSearch/LEAP algorithm \cite{smith_leap_2021}.
The original QGo algorithm was re-implemented using the partitioning and synthesis algorithms provided by the \emph{BQSKit} toolkit \cite{bqskit}.

\begin{figure*}
    \begin{subfigure}{\textwidth}
        \centering
        \includegraphics[width=.9\textwidth]{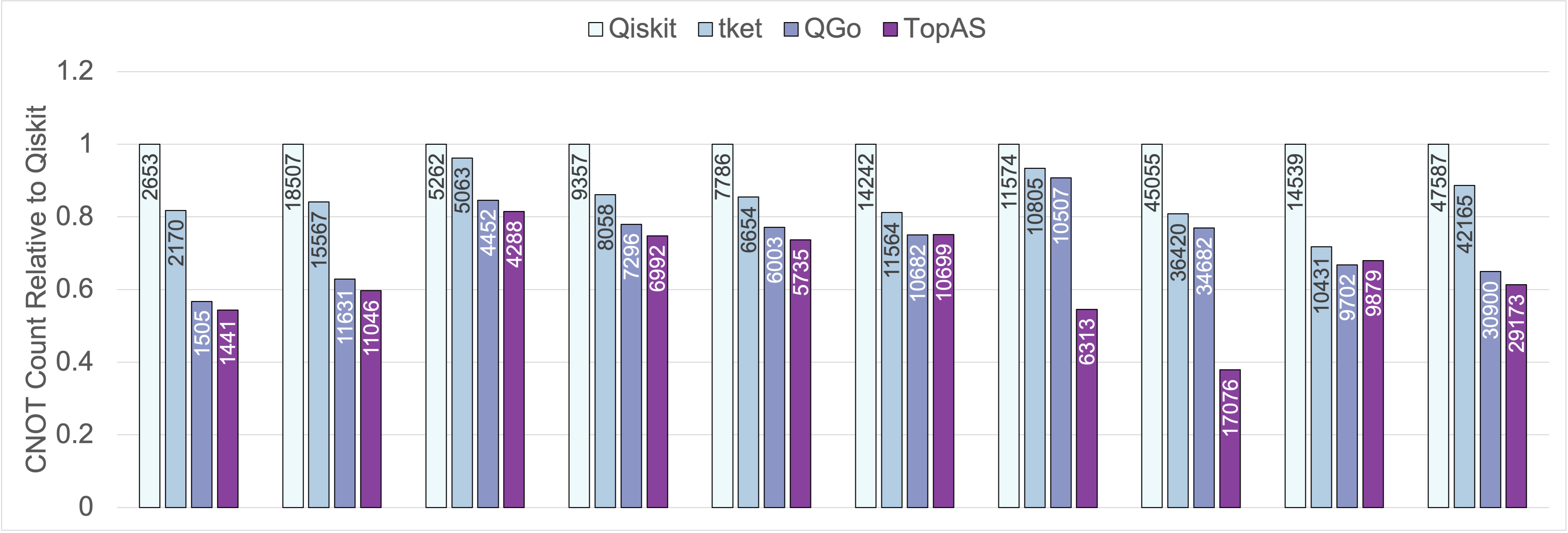}
    \end{subfigure}
    \begin{subfigure}{\textwidth}
        \centering
        \includegraphics[width=.9\textwidth]{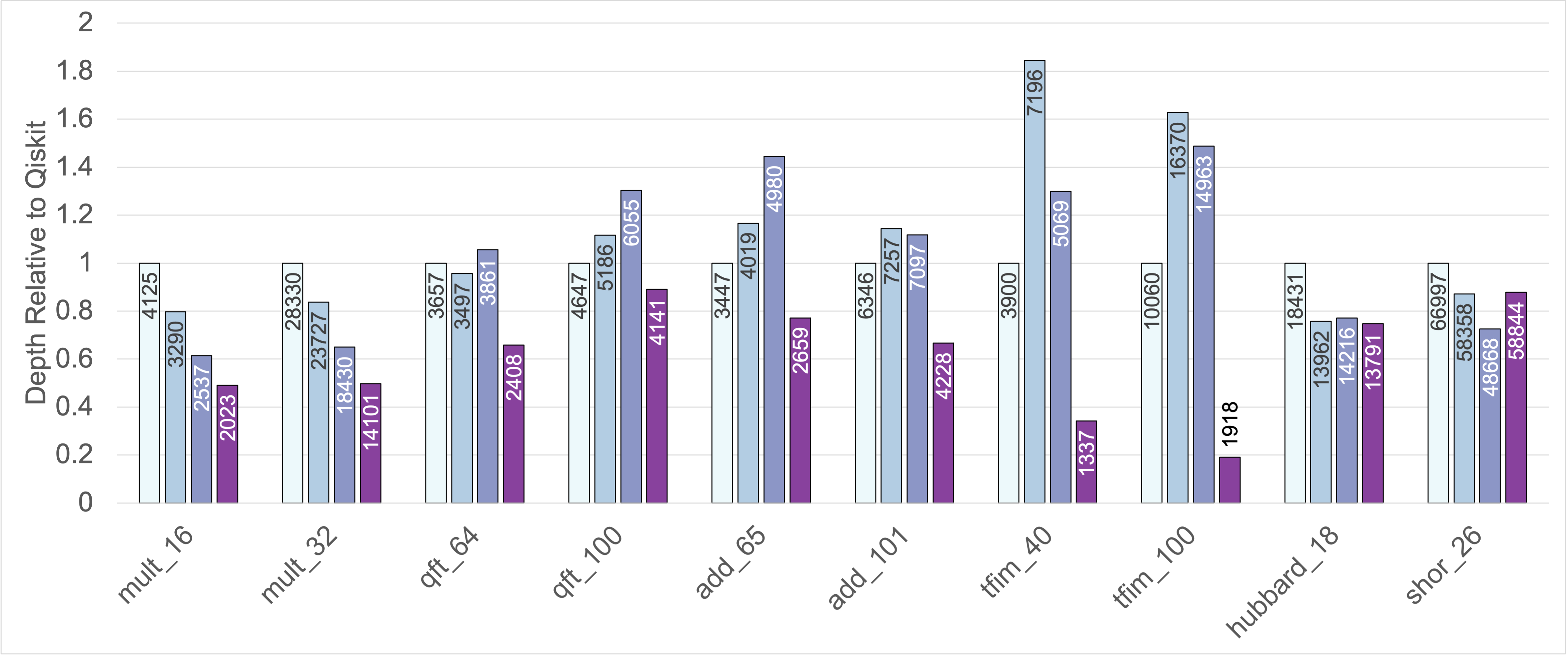}
    \end{subfigure}
    \caption{
        Comparison of relative CNOT count (top) and depth (bottom) for circuits mapped to the Google style mesh physical topology.
        CNOT count and depth is shown relative to optimized circuits mapped using \emph{Qiskit}'s SABRE Swap algorithm.
        Lower depth corresponds directly to better program runtimes on machines with parallel gate execution, lower depth and CNOT count correspond to improved execution probability on noisy machines.
    }
    \label{fig:main_mesh}
\end{figure*}

\begin{figure*}
    \begin{subfigure}{\textwidth}
        \centering
        \includegraphics[width=.9\textwidth]{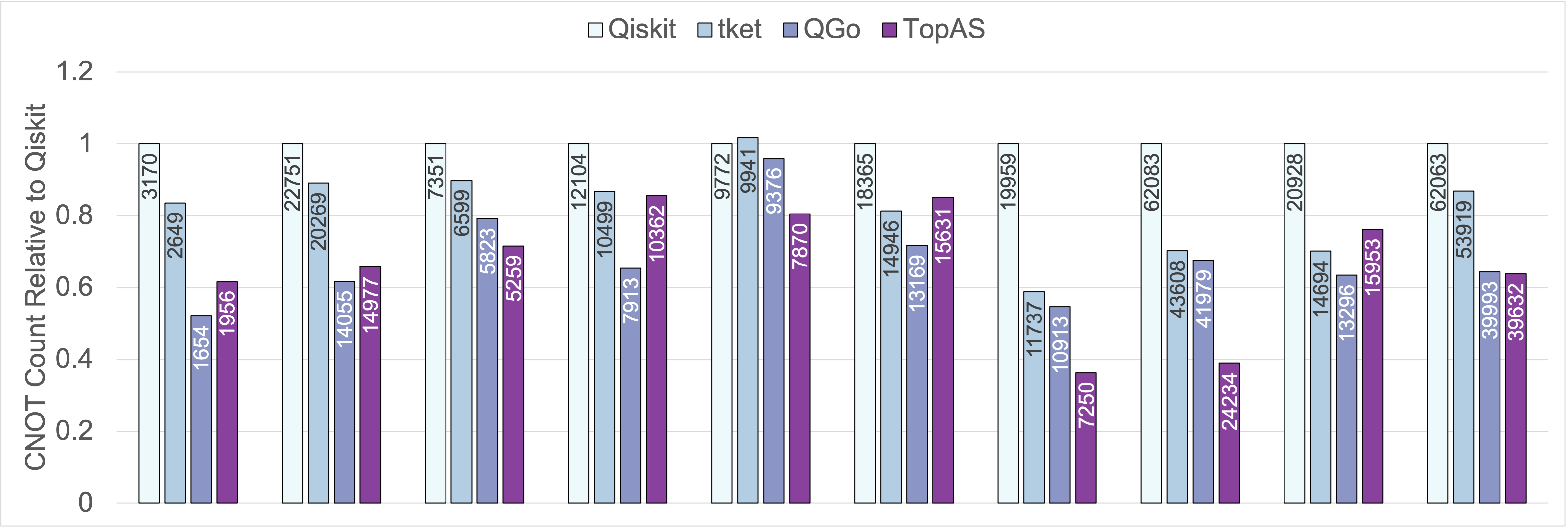}
    \end{subfigure}
    \begin{subfigure}{\textwidth}
        \centering
        \includegraphics[width=.9\textwidth]{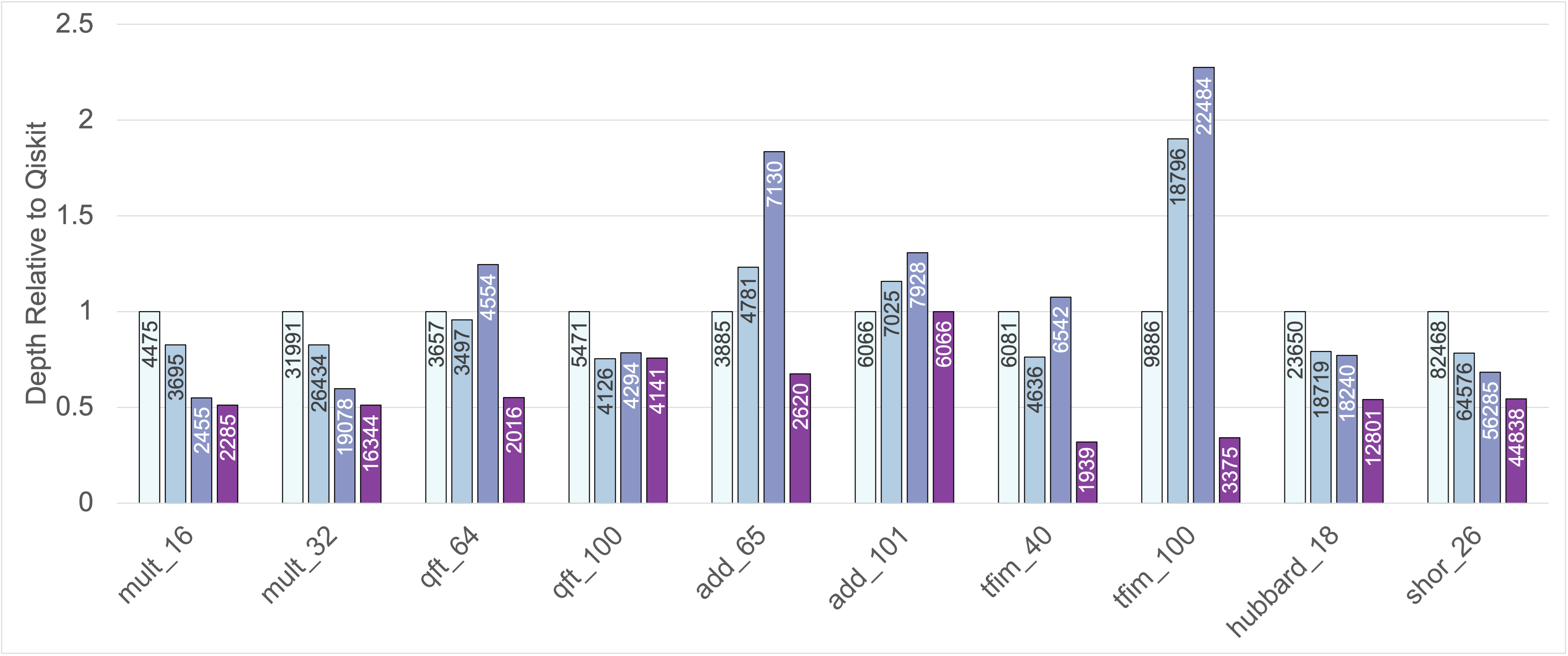}
    \end{subfigure}
    \caption{
        Comparison of relative CNOT count (top) and depth (bottom) for circuits mapped to the IBM style falcon physical topology.
        CNOT count and depth is shown relative to optimized circuits mapped using \emph{Qiskit}'s SABRE Swap algorithm.
        Lower depth corresponds directly to better program runtimes on machines with parallel gate execution, lower depth and CNOT count correspond to improved execution probability on noisy machines.
    }
    \label{fig:main_falcon}
\end{figure*}

Performance was evaluated using a variety of wide quantum circuit benchmarks.
The SupermarQ benchmark suite \cite{tomesh_2022_supermarq} proposed a \emph{volume} metric to describe a benchmark suite's diversity.
Our selection of benchmarks has a SupermarQ volume of $7.46 \times 10^{-6}$. 
This small selection of circuits outperforms many other benchmark suites in SupermarQ volume due to its focus on real algorithms.
Our suite's SupermarQ volume score is limited by the fact that none of our benchmarks include intermediate measurements. 
This decision was made as the primary focus in the evaluation of synthesis based optimization tools is purely depth and multi-qubit gate count reduction. 
Each benchmark assumes qubits are measured after the final program gate is executed.

The \emph{qft} and \emph{shor} circuits were generated using \emph{Qiskit}. 
The \emph{add} and \emph{mult} circuits were generated with \cite{hassan_qarithmetic_2021}.
The \emph{hubbard} benchmark was generated with help from \cite{mcclean_openfermion_2019}.
TFIM circuits were generated using the ArQtic tool \cite{bassman_arqtic_2021}.
Benchmarks are labeled so that the number following the benchmark name indicates the number of qubits in the circuit.
These benchmarks were selected as they were found to be relatively easily generated, and represent a variety of wide circuits that may soon be executable on quantum machines.
Results for the QGo and TopAS tools were collected using NERSC's \emph{Perlmutter} supercomputer.

Figures \ref{fig:main_mesh} and \ref{fig:main_falcon} show the relative CNOT gate count and depth of circuits optimized and mapped to the mesh and falcon physical topologies illustrated in Figure \ref{fig:physical_topologies}.
The CNOT gate count and depth are normalized by the results obtained for circuits optimized then mapped using \emph{Qiskit}'s SABRE mapping algorithms.
A synthesis threshold distance of $\epsilon = 1\times10^{-10}$ and a partition width of $k=4$ was used for both QGo and TopAS.

On average, the TopAS tool is able to reduce CNOT count compared to \emph{Qiskit} and $t|ket\rangle$ by 35.9\% and 24.3\% respectively when targeting a mesh physical topology. 
For the falcon physical topology, TopAS outperforms \emph{Qiskit} and $t|ket\rangle$ by an average of 33.4\% and 19.1\% respectively.
For the \emph{tfim\_100} circuit, TopAS outperforms \emph{Qiskit} by 62.1\% and $t|ket\rangle$ by 53.1\%. 
For the \emph{shor\_26} benchmark, TopAS outperforms \emph{Qiskit} by 38.7\% and $t|ket\rangle$ by 30.8\%. 
Reductions in circuit depth are even greater, averaging 38.6\% and 37.6\% lower compared to \emph{Qiskit} and $t|ket\rangle$ across all benchmarks.

In most cases, the TopAS tool is able to produce circuits with fewer CNOTs and much lower depth than the QGo tool, especially for the mesh physical topology.
On average, TopAS outperforms QGo by 11.5\% in CNOT count and 34.3\% in depth when targeting the mesh physical topology.
For the falcon physical topology, TopAS only outperforms QGo by an average of 0.8\% in CNOT count, but maintains a 37.3\% average depth reduction.

QGo optimized circuits always reduce CNOT count compared to unsynthesized input circuits, but sometimes lead to deeper circuits.
This is because when synthesizing pure SWAP gates, the QSearch algorithm tends to find the more expensive depth 5 implementation illustrated in Figure \ref{fig:swap_gate}.
TopAS is able to avoid this pitfall by synthesizing partitions that do not contain SWAP gates, thus resulting in much reduced circuit depths.

As discussed in Section \ref{section:synth_logical_quantum_circs}, working on the logical, unmapped, quantum circuit also allows TopAS to form fewer partitions compared to QGo.
This results lower upper bounds on the amount of synthesis error for each benchmark.
Synthesis error upper bounds are given in Table \ref{table:error_and_avg_cnots}.
This table also lists the average number of CNOTs per partition for all benchmarks.

TopAS is able to outperform QGo in CNOT count for all benchmarks when targeting the mesh physical topology except for the \emph{hubbard\_18} circuit benchmark.
The discrepancy in performance for this benchmark is primarily explained by the average size of partitions formed.
Table \ref{table:error_and_avg_cnots} shows that for this circuit, TopAS is only able to form partitions with an average of 3.7 CNOT gates.
Figure \ref{fig:partition_summary} illustrates that typically, partitions with this number of CNOTs tend not to reduce.
Partitions formed by QGo tend to include SWAP gates, which greatly increases the number of CNOTs per partition (6.5 CNOT gates on average for \emph{hubbard\_18}).
For the \emph{hubbard\_18} circuit, TopAS's logical partitioning was simply not able to form few enough partitions to compensate for the small average number of CNOT gates per partition.
Using a larger partition width, or a partitioning algorithm that is able to produce larger partitions on average would likely improve the performance of TopAS for the \emph{hubbard} benchmark.
This point also partially explains the drop in TopAS' performance compared to QGo for benchmarks targeting the falcon physical topology.

\begin{figure}
    \begin{subfigure}[t]{0.48\textwidth}
        \centering
        \includegraphics[width=\textwidth]{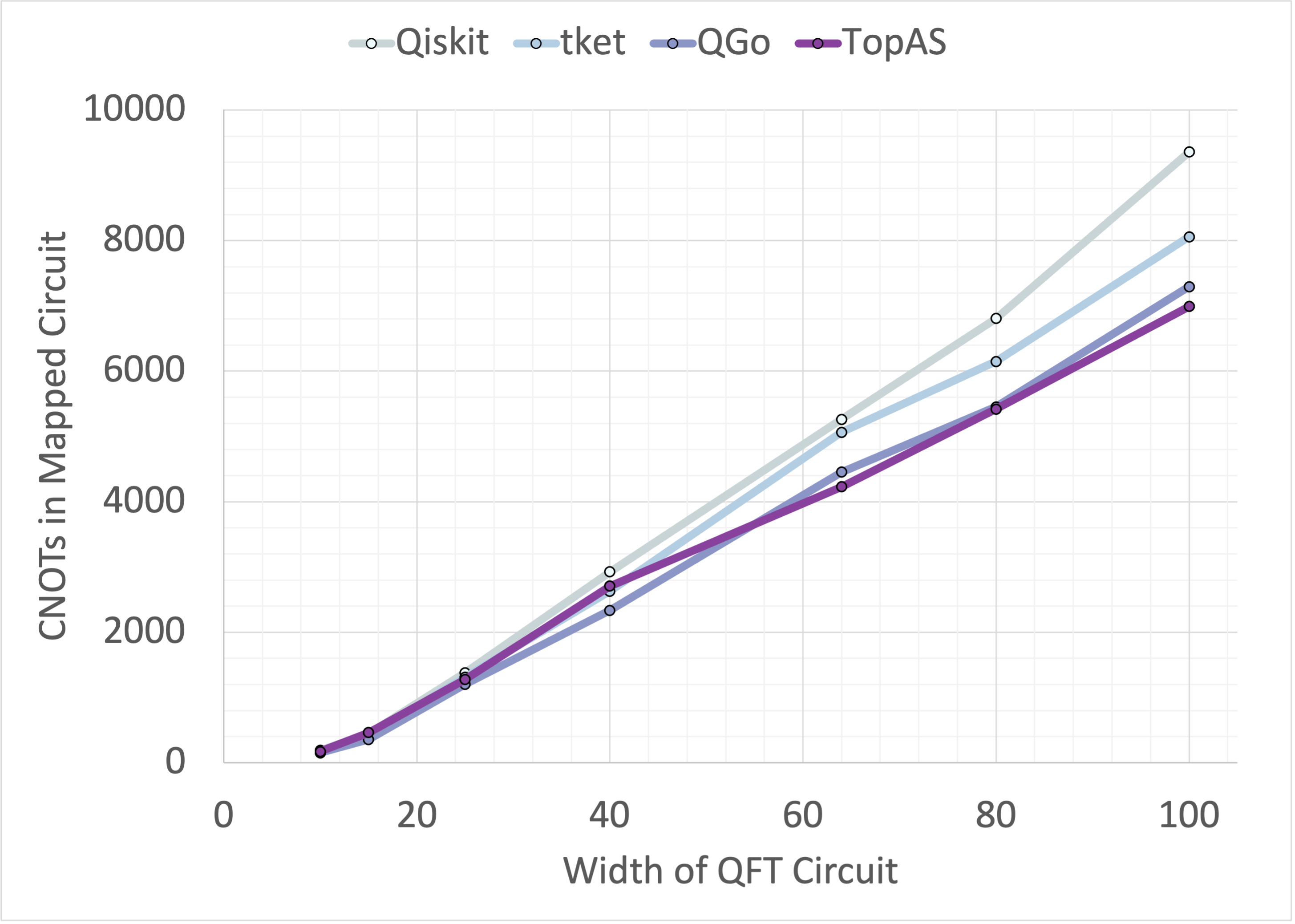}
        \label{fig:qft_scaling}
    \end{subfigure}
    \begin{subfigure}[t]{0.48\textwidth}
        \centering
        \includegraphics[width=\textwidth]{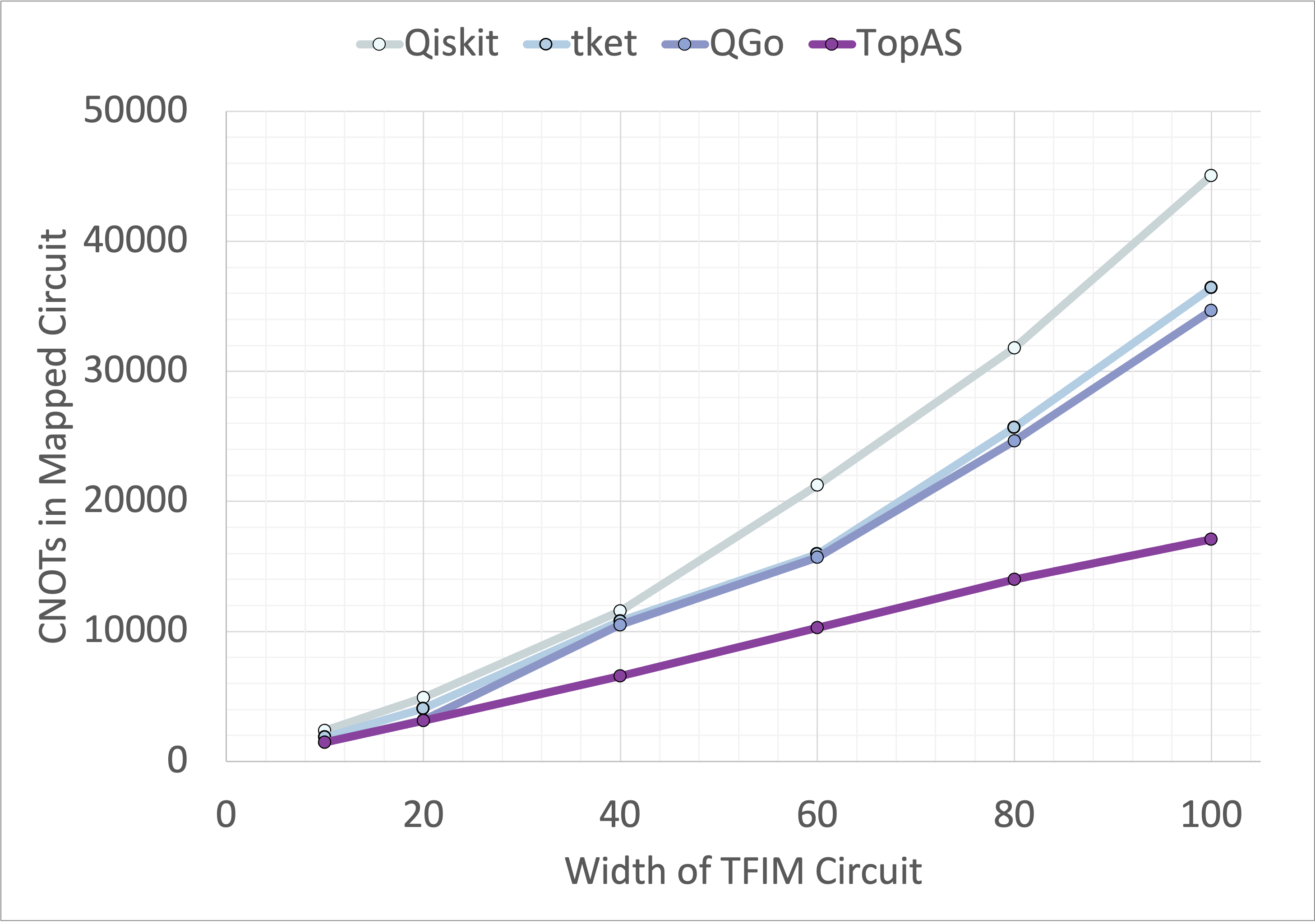}
        \label{fig:tfim_scaling}
    \end{subfigure}
    \caption{
        Number of CNOTs in QFT (top) and TFIM (bottom) circuits mapped to the 2D mesh physical topology as a function of circuit width.
    }
    \label{fig:scaling}
\end{figure}

The scalability of the TopAS tool compared to \emph{Qiskit}, $t|ket\rangle$, and QGo is illustrated in Figure \ref{fig:scaling}.
Each tool was used to optimize and map QFT and TFIM circuits with widths of 10-100 qubits.
Despite performing poorly in comparison at small circuit widths, TopAS maintains an advantage compared to other optimization tools as the circuit width increases.
In the case of certain circuits such as the TFIM circuits, TopAS' advantage increases with circuit width.

\section{Discussion}
\label{section:discussion}
TopAS is able to produce circuits with fewer multi-qubit gates and lower depth by partitioning logical quantum circuits and matching subcircuits with sparse qubit subtopologies in such a way that balances the demands of synthesis and mapping algorithms.
Partitioning and using unitary synthesis to optimize the logical quantum circuit allows for smaller circuits to be passed to mapping algorithms.
In the case of the \emph{tfim\_40} and \emph{tfim\_100} benchmarks, the logical connectivity graphs are simply linear path graphs of order 40 and 100 respectively.
These circuits have implementations that do not require any SWAP operations, as they are directly embeddable within the mesh physical topology.
However, mapping algorithms implemented in \emph{Qiskit} and $t|ket\rangle$ are unable to find these placements, likely due to the large widths and depths of these circuits.
By optimizing the logical quantum circuits, TopAS is able shorten circuits in such a way that allows for these mapping algorithms to find better placements.
This advantage in mapping performance is maintained in the falcon topology case, where the TFIM circuits are no longer directly embeddable in the physical topology.

Using synthesis to map partitions to restrictive subtopologies also improves the performance of the mapping algorithm.
Mapping to subtopologies that are easily embedded within the physical topology means that fewer SWAPs are needed within the execution time of partitions.
This effect is illustrated in Figure \ref{fig:internal_swaps}.
The \emph{neighbor aware} subtopology selection mechanism illustrated in Figure \ref{fig:neighbor_aware} helps to maximize the reuse of edges in the physical topology between the execution of partitions.
The combination of these design choices allows for TopAS to outperform other optimization and mapping tools.

Section \ref{section:data} demonstrates how TopAS is able to reduce both CNOT count and circuit depth significantly compared to other tools.
These metrics both play a major role in the likelihood that a circuit will be executed correctly.
The amount of error due to synthesis (shown in Table \ref{table:error_and_avg_cnots}) is far lower than that introduced by gate noise and decoherence.
In total, optimizing wide circuits with TopAS therefore greatly increases circuit fidelity compared to other tools.

Because mapping is done after partitioning and synthesis in the TopAS program flow, it is not limited to using a single mapping algorithm.
We observed that although TopAS synthesizes partitions to subtopologies embedded within the physical topology, it is often the case that the mapping algorithm disrupts the execution of synthesized subcircuits.
When partitions are executed atomically, there is a solution to the routing problem such that no SWAP gates are needed during the partition's execution.
A mapping algorithm that is aware of higher level structures than the primitive gate set may therefore further improve the performance of TopAS.
Such a tool was put forth by the authors of \cite{Duckering_2021}, but it is only able to consider Toffoli gates instead of arbitrary $k$ qubit unitary operations.
A full \emph{partition aware} mapping algorithm is therefore likely necessary to reap the full benefits of this strategy.

Although TopAS' set of valid synthesis subtopologies only includes graphs that are embedded within the physical topology, it does not consider whether a set of subtopologies can be packed into the physical topology simultaneously.
This fact explains the discrepancy in TopAS' performance between the mesh and the falcon physical topologies.
For example, although the star subtopology is embedded within the falcon physical topology, far fewer star graphs can be packed into the falcon topology than the mesh topology.
A subtopology selection process that more precisely weighs the frequency with which subgraphs appear in a physical topology would thus likely further improve the performance of TopAS.

The two primary factors that limit the scalability of TopAS are the runtime of the partitioning algorithm used and the runtime of the synthesis algorithm used.
In the case where qubits in the logical circuit interact with all other qubits, the runtime of the \emph{scan} partitioner is $O {n \choose k}$, where $n$ is the circuit width and $k$ is the partition width.
This poor scaling limited the width of benchmark circuits tested to 100 qubits.
Other partitioning schemes capable of forming large partitions on average are therefore needed to optimize larger quantum circuits with TopAS.
Algorithms such as QFAST \cite{younis_qfast_2020} effectively increase the width of circuits that can be optimized, but in our experience tend not to perform as well as QSearch/LEAP in the range of 1-25 CNOT gates.
Increasing the partition width tends to increase the number of CNOTs per partition, which may provide more opportunity for improvement.
Increasing the partition width also has the effect of adding more graphs to the set of valid synthesis subtopologies.
Supposing that the optimization algorithm scales, this may further improve the performance of TopAS by allowing for a greater amount of mapping to be handled by the optimization algorithm.

As shown in Figure \ref{fig:scaling}, the number of CNOTs in QFT circuits optimized with the TopAS tool grows roughly linearly with the number of qubits in the range shown.
As the QFT is a common component of more complex quantum algorithms, the scaling of this circuit is of particular importance.
With a width of 100 qubits, the TopAS optimized QFT circuit contains 23.9\% and 11.7\% fewer CNOTs than the \emph{Qiskit} and $t|ket\rangle$ optimized and mapped circuits.
Notably, the rate of increase in CNOTs per qubit in the QFT circuit seems to grow more slowly for TopAS than for \emph{Qiskit} and $t|ket\rangle$.
Because QGo only optimizes circuits that have already been mapped by these tools, we expect that TopAS' performance advantage over QGo can improve at larger (100-1000 qubit) circuit widths.
This effect can clearly be seen in the TFIM circuit scaling results.

Although out of the scope for the current paper, an obvious next question would be to see what happens when TopAS is combined with QGo.
We have clearly shown that a single application of TopAS' logical circuit synthesis almost always outperforms a single round of QGo optimization.
When applying both TopAS and QGo to the \emph{hubbard\_18} benchmark mapped to the 2D mesh physical topology (a benchmark which TopAS fails to reduce CNOT count more than QGo for the mesh topology), CNOT count was reduced by approximately 5\% compared to QGo alone.
Thus, combining multiple levels of optimization has the potential to further improve the performance of quantum circuits.

\section{Conclusion}
\label{section:conclusion}
In this work, we have presented TopAS, a topology aware synthesis tool that optimizes wide quantum circuits.
By optimizing quantum circuits using unitary synthesis before they are mapped to restrictive qubit topologies, TopAS preconditions circuits so that they are minimized by synthesis and made easier to route by mapping algorithms.
TopAS is able to reduce CNOT count and circuit depth, and thus improves circuit performance, compared to other state of the art synthesis based optimization tools targeting wide circuit optimization.
TopAS also outperforms the optimization and mapping frameworks provided by \emph{Qiskit} and $t|ket\rangle$.
Because CNOT count and circuit depth are significantly reduced, the likelihood that circuits optimized with TopAS execute successfully on noisy, realistic, near term qubit topologies is greatly increased.
Further reduction is possible on TopAS optimized circuits by applying successive rounds of post-mapping circuit optimization.

\section*{Acknowledgements}
\label{section:acknowledgements}
This work was supported by the DOE under contract DE-5AC02-05CH11231 through the Office of Advanced Scientific Computing Research (ASCR) Quantum Algorithms Team and Accelerated Research in Quantum Computing programs, and by the NSF Challenge Institute for Quantum Computation (CIQC) program under award OMA-2016245.


\bibliographystyle{IEEEtranS}
\bibliography{refs}

\end{document}